\documentclass[11pt,dvips,twoside,letterpaper]{article} 
\usepackage{times,fancyhdr}
\usepackage{graphics}
\usepackage{geometry}
\usepackage{titlesec}


\makeatletter 
\def\cleardoublepage{\clearpage\if@twoside \ifodd\c@page\else%
    \hbox{}%
    \thispagestyle{empty}
    \newpage%
    \if@twocolumn\hbox{}\newpage\fi\fi\fi} 
\makeatother 

\titlelabel{\thetitle.\quad}

\def\figurename{Figure} 
\makeatletter
\renewcommand{\fnum@figure}[1]{\figurename~\thefigure.}
\makeatother

\def\tablename{Table} 
\makeatletter
\renewcommand{\fnum@table}[1]{\centering\bfseries{\tablename~\thetable.}}
\makeatother

\usepackage{epsfig}
\usepackage{subfigure}
\usepackage{xspace}

\newcommand{\ra}{\textsc{r.a.}\xspace}

\newcommand{\be}{\begin{equation}}
\newcommand{\ee}{\end{equation}}
\newcommand{\ba}{\begin{eqnarray}}
\newcommand{\ea}{\end{eqnarray}}

\newcommand{\bfi}{\begin{figure}
\epsfxsize=9cm
\epsffile}
\newcommand{\efi}{\end{figure}}
\newcommand{\bi}{\begin{itemize}}
\newcommand{\ei}{\end{itemize}}

\newcommand{\etal}{\emph{et al.}}

\begin{document}
\title{
{\begin{flushleft}
\vskip 0.45in
{\normalsize\bfseries\textit{Chapter~xx}}
\end{flushleft}
\vskip 0.45in
\bfseries\scshape On the Observation of the Cosmic Ray Anisotropy below 10$^{15}$ eV}}
\author{\bfseries\itshape G. Di Sciascio$^{1}$ and R. Iuppa$^{1,2}$\\
$^{1}$INFN, sez. di Roma Tor Vergata \\
$^{2}$Dipartimento di Fisica dell'Universit\'a degli studi di Roma Tor Vergata}
\date{}
\maketitle
\thispagestyle{empty}
\setcounter{page}{1}

%
%
%

\begin{abstract}
The measurement of the anisotropy in the arrival direction of cosmic rays is complementary to the study of their energy spectrum and chemical composition to understand their origin and propagation. It is also a tool to probe the structure of the magnetic fields through which cosmic rays travel.

As cosmic rays are mostly charged nuclei, their trajectories are deflected by the action of galactic magnetic field they propagate through before reaching the Earth atmosphere, so that their detection carries directional information only up to distances as large as their gyro-radius. If cosmic rays below $10^{15}{\rm\,eV}$ are considered and the local galactic magnetic field ($\sim3{\rm\,\mu G}$) is accounted for, gyro-radii are so short that isotropy is expected. At most, a weak di-polar distribution may exist, reflecting the contribution of the closest CR sources.

However, a number of experiments observed an energy-dependent \emph{``large scale''} anisotropy in the sidereal time frame with an amplitude of about 10$^{-4}$ - 10$^{-3}$, revealing the existence of two distinct broad regions: an excess distributed around 40$^{\circ}$ to 90$^{\circ}$ in Right Ascension (commonly referred to as ``tail.in'' excess) and a deficit (the ``loss cone'') around 150$^{\circ}$ to 240$^{\circ}$ in Right Ascension.
In recent years the Milagro and ARGO-YBJ collaborations reported the of a ``medium'' scale anisotropy inside the tail-in region. The observation of such small features has been recently claimed by the IceCube experiment also in the Southern hemisphere.

So far, no theory of cosmic rays in the Galaxy exists which is able to explain the origin of these different anisotropies leaving the standard model of cosmic rays and that of the galactic magnetic field unchanged at the same time.

Although observations of the CR anisotropy were reported up to 10$^{20}$ eV, this work is focused on the energy range below $10^{15}{\rm\,eV}$, where the counting rates of the detectors are high enough and the evidence for the anisotropy well established.
\end{abstract}
%
\section{Introduction}
Cosmic rays (CRs) are the most outstanding example of accelerated particles. Understanding their origin and propagation through the interstellar medium is a fundamental problem which has a major impact on models of the structure and nature of the Universe.

The observed primary CR energy spectrum exceeds 10$^{20}$ eV showing a few basic characteristics: (a) a power-law behaviour $\sim$ E$^{-2.7}$ until the so-called ``knee'', a small downwards bend around few PeV; (b) a power-law behaviour $\sim$ E$^{-3.1}$ beyond the knee, with a slight dip near 10$^{17}$ eV, sometimes referred to as the ``second knee''; (c) a transition back to a power-law $\sim$ E$^{-2.7}$ (the ``ankle'') around $10^{18}$~eV; (d) a cutoff due to extra-galactic CR interactions with the Cosmic Microwave Background (CMB) around 10$^{20}$ eV (the Greisen-Zatsepin-Kuzmin effect).
All these features are believed to carry fundamental information to shed light on the key question of the CR origin.

CRs below 10$^{17}$ eV are expected to be mainly galactic, produced and accelerated in SuperNova (SN) blast waves, which can provide the necessary energy input and naturally produce particles with a power law energy spectrum and spectral index close to the value inferred from the observations.
The SN acceleration mechanisms are limited by the rate at which the particles can gain energy and by the lifetime of the shock.
As an example, for a typical SN, Lagage and Cesarsky \cite{lagage} calculated that this limit corresponds, for a particle of charge $Z$ (in natural units), to a total energy of $\sim Z\times$ 10$^{14}$ eV.

\pagestyle{fancy}
\fancyhead{}
\fancyhead[EC]{G. Di Sciascio and R. Iuppa}
\fancyhead[EL,OR]{\thepage}
\fancyhead[OC]{On the Observation of the Cosmic Ray Anisotropy below 10$^{15}$ eV}
\fancyfoot{}
\renewcommand\headrulewidth{0.5pt} 
As CRs are mostly charged nuclei, their paths are deflected and highly isotropized by the action of galactic magnetic field (GMF) they propagate through before reaching the Earth atmosphere. The GMF is the superposition of regular field lines and chaotic contributions. Although the strength of the non-regular component is still under debate, the local total intensity is supposed to be $B=2\div 4\textrm{ $\mu$G}$ \cite{beck01}. In such a field, the gyro-radius of CRs is given by ${r}_{a.u.}\approx 100_{\textrm{\scriptsize{TV}}}$, where $r _{a.u.}$ is in astronomic units and R$_{\textrm{\scriptsize{TV}}}$ is the rigidity in TeraVolt. 
Clearly, there is very little chance of observing a point-like signal from any radiation source below $10^{17}{\rm eV}$, as they are known to be at least several hundreds parsecs away.

If it is true that magnetic fields are the most important ``isotropizing'' factor when they randomly vary on short distances, it is clear as much that some particular features of the magnetic field at the boundary of the solar system or farther might focus CRs along certain lines and the observed arrival direction distribution turns out to be consequently an-isotropic.

As it will be discussed in this paper, different experiments observed an energy-dependent \emph{``large scale''} anisotropy in the sidereal time frame with amplitude spanning 10$^{-4}$ to 10$^{-3}$, from tens GeV to hundreds TeV, suggesting the existence of two distinct broad regions, one showing an excess of CRs (named \emph{``tail-in''}), distributed around 40$^{\circ}$ to 90$^{\circ}$ in Right Ascension (R.A.). The other a deficit (the so-called \emph{``loss cone''}), distributed around 150$^{\circ}$ to 240$^{\circ}$ in R.A..

The origin of the galactic CR anisotropy is still unknown, but the study of its evolution over the energy spectrum has an important valence to understand the propagation mechanisms and the structure of the magnetic fields through which CRs have traveled.

The propagation in the Galaxy of CRs trapped by the magnetic field is usually described in terms of diffusion at least up to 10$^{16-17}$ eV \cite{berez90}.
The gyro-radius of particles is smaller than the largest scale (100 - 300 pc) on which the turbulent component of the GMF is believed to vary. Therefore, the CR propagation resembles a random walk. The diffusion produces density gradients and thus an anisotropy as intense as $\delta$ takes place: $\delta$ = $\frac{3D}{v} \frac{\nabla N}{N}$, where $v$ is the particle velocity, $N$ the particle density and $D$ the diffusion coefficient expected to increase with the rigidity $R$.
The models proposed to describe the CR propagation adopt different dependence of $D$ with energy: $D\propto R^{0.6}\ \rm{cm}^2\ \rm{ s}^{-1}$ in the ``leaky box'' approximation with a regular magnetic field \cite{swordy95}; $D\propto R^{0.3}\ \rm{ cm}^2\ \rm{ s}^{-1}$, for models with re-acceleration and an additional random turbulent component of the GMF; $D\propto R^{0.15-0.20}\ \rm{ cm}^2\ \rm{s}^{-1}$ for models with a Hall diffusion \cite{ptuskin93}. A measurement of the evolution of anisotropy with the energy is therefore important to disentangle between different models of CR propagation in the Galaxy.

Moreover, the study of the anisotropy can clarify the key problem of galactic CRs, i.e. the origin of the knee. Indeed, if the knee is due to an increasing inefficiency in CR containment in the Galaxy a change in $\delta$ is expected. If, on the contrary, the knee is related to the inefficiency of the acceleration mechanism, we do not expect a change of anisotropy when nearing the knee.

Around 10$^{18}$ eV the gyro-radius of CRs becomes comparable to the thickness of the galactic disk, therefore their confinement in the Galaxy is less effective and the diffusive motion loses importance. As a consequence, if the CR sources are distributed mainly in the galactic plane, an increase of the anisotropy toward the galactic disk is expected, due to the proton component. 

Above 10$^{18}$ eV, the CRs are believed to be extra-galactic. With increasing energy the \emph{``charged-particle astronomy''} becomes possible in principle, because the magnetic field should not affect significantly the CR propagation.
Nevertheless, the uncertainty about the extra-galactic magnetic field structures and about the CR chemical composition at that energy makes challenging to achieve significant results.
Although observations of anisotropy were reported up to energy around 10$^{20}$ eV, in this paper we review measurements and models concerning CRs below $10^{15}\ \rm{eV}$, where the counting rate of detectors is high enough to give solid arguments on which to ground the discussion.

The paper is organized as follows. In the section \ref{sec:methods} the detection methods used so far are summarized. The existing data on the CR anisotropy below the knee are reviewed in the section \ref{sec:datareview}. In the section \ref{sec:models} the different models proposed to interpret data are reported. Some conclusions follow.
\section{Methods of Detection}
\label{sec:methods}
The amplitude of the anisotropy is so small ($\approx 10^{-3}$ or less) that the experimental approach to detect it has to be sensitive on necessity. In order to measure such a tiny effect, large exposures are needed, i.e. large instrumented areas and long-lasting data acquisition campaigns. Until now, only ground-based detectors demonstrated to have the required sensitivity, making use of the Earth's rotation to order data in solar and sidereal time.

As the Earth rotates, the field of view of ground-based detectors points towards different directions at different times, sweeping out a cone of constant declination. 

The CR arrival direction is given after the definition of a reference frame. If Sun-related effects are looked for, a system of coordinates co-moving with the Earth has to be used. Usually, it is a spherical system centered on the Earth. Instead, if phenomena originated outside the heliosphere are considered, a reference frame co-moving with the Sun is set, usually a spherical system centered there\footnote{Or on the Earth, that is the same because of the distances under consideration.}. Whatever the spherical system is, because of the Earth rotation, the direction $\phi$ is periodically observed from ground-based detectors, that is why the $\phi$ axis is regarded as a time axis. The ``solar'' or ``universal'' time is used when the Sun is wanted to be at rest in the reference frame, whereas the ``sidereal'' time is used when the Galactic center is wanted to be at rest.

The \emph{sidereal anisotropy} is spoken about when the sidereal time is used, whereas the \emph{solar anisotropy} is measured by using the solar time. The \emph{anti-sidereal} time analysis is often performed too, as no signal is expected in this non-physical reference frame and any positive result there can be used as estimation of the systematic uncertainty on the sidereal time detection.

Once the reference frame is defined, data are collected and ordered accordingly. There are a number of ways in which the degree of anisotropy of a given distribution can be defined. Perhaps the most traditional one in CR physics is:
\begin{equation}
  \delta=\frac{I_{max}-I_{min}}{I_{max}+I_{min}}
  \label{eq:delta_anisotropy}
\end{equation}
where $I_{max}$ and $I_{min}$ are the maximum and the minimum observed intensity. This definition is the most general to be given, as no hypothesis on the form of the anisotropy is needed: either it is a peak in a smooth distribution or a di-polar modulation, the definition (\ref{eq:delta_anisotropy}) can be applied. 

In 1975 Linsley proposed to perform analyses in right ascension only, through the so-called ``harmonic analysis'' of the counting rate within a defined declination band, given by the field of view of the detector \cite{linsley75}.
In general, one calculate the first and second harmonic from data by measuring the counting rate as a function of the sidereal time (or right ascension), and fitting the result to a sine wave. The Rayleigh formalism allows to evaluate the amplitude of the different harmonics, the corresponding phase (the hour angle of the maximum intensity) and the probability for detecting a spurious amplitude due to fluctuations of a uniform distribution.

Let $\alpha_1,\alpha_2,\dots\alpha_n$  be the right-ascension of the $n$ collected events. From this data series, the experimenter has to determine the components of the two-dimensional vector amplitude (or, equivalently, the scalar amplitude $r$ and phase $\phi$).

The \ra distribution $f(\alpha)$ can be represented with a Fourier series:
\begin{equation}
f(\alpha) =\frac{a_0}{2}+ \sum_{k=1}^{\infty} (a_k \cos k\alpha + b_k \sin k\alpha )= \frac{a_0}{2} + \sum_{k=1}^\infty r_k \sin (k\alpha + \phi_k)
\end{equation}
where $a_k=r_k \sin\phi_k$ and $b_k=r_k \cos\phi_k$ are the Fourier coefficients and 
\begin{equation}
r_k = \sqrt{a_k^2 + b_k^2}
\end{equation}
\begin{equation}
\phi_k=\tan\frac{a_k}{b_k}
\end{equation}
are the amplitude and the phase of the $k^{th}$ harmonic, respectively. It is known that the  Fourier coefficients can be computed from the $\alpha_m$ data series by means of the equations:
\begin{equation}
a_k = \frac{2}{n}\sum_{i=1}^n \cos k\alpha_i
\end{equation}
\begin{equation}
b_k = \frac{2}{n}\sum_{i=1}^n \sin k\alpha_i
\end{equation}
where $n$ is the number of data points.
This formalism can be applied up to whichever order $k$, but historically experiments never had the sensitivity to go beyond $k=3$ and anthologies are usually compiled about $k=1$ only.
If a non-zero first harmonic amplitude is found, it is important to estimate the chance probability $P$ that it is a fluctuation of an isotropic distribution.
If the sample $\alpha_1$, $\alpha_2$, ..., $\alpha_n$ is randomly distributed between 0 and 2$\pi$, then, as $n\rightarrow \infty$, the probability $P$ of obtaining an amplitude greater than or equal to $r$ is given by the well-known Rayleigh formula  
\begin{equation}
P(\geq r) = \exp(-k_0)
\end{equation}
where $k_0 = (nr^2)/4$.

In general, harmonic analysis is effective for revealing broad directional features in large samples of events measured with poor angular resolution but good stability.

The technique is rather simple: the greatest difficulty lies in the treatment of the data, that is, of the counting rate themselves. In fact, the expected amplitudes are very small with related statistical problems: long term observations and large collecting areas are required.
Spurious effects must be kept as low as possible: uniform detector performance both over instrumented area and over time are necessary as well as operational stability. In addition, CR experiments typically suffer from large variations of atmospheric parameters as temperature and pressure, which translate into changes of the effective atmospheric depth, affecting the CR arrival rate.

Therefore, the measured rate must be corrected precisely for instrumental and atmospheric effects to prevent any misinterpretation of CR flux variations.

The \emph{East-West method} was designed to avoid performing such corrections, preventing potential subsequent systematic errors introduced by data analysis.
The original idea was proposed in the early 1940s to study asymmetries in the flux of solar CRs \cite{kolhoster41,alfver43,elliot51} and was later applied by Nagashima to extensive air showers (EASs) at higher energy \cite{nagashima89}. It is a differential method, as it is based on the analysis of the difference of the counting rates in the East and West directions \cite{bonino11}. 
By considering this, the method is able to eliminate any atmospheric effect or detector bias producing a common variation in both data groups.
\vspace{3mm}

For primary CRs with energy lower than 10$^{12}$ - 10$^{13}$ eV, underground muon detectors have been widely used. 
The muon rate deep underground can be affected by a spurious  periodic modulation as the result of the competition between pion decay and interaction in the upper atmosphere. As the atmosphere cools at night (or during the winter), the density increases and the pions produced in CR collisions with the atmosphere are fractionally more likely to interact than decay when compared with the day (or summer) when the climate it is warmer. The daily modulation introduced by solar heating and cooling is seen when muons are binned in solar-diurnal time; yearly or seasonal modulations are seen when events are binned with a period of a (tropical) year \cite{macro97,macro02}.
With increasing energy an increasing fraction of high energy secondary pions interact in the higher atmosphere instead of decaying to muons and the muon intensities decrease too rapidly to yield sufficient statistics for high-energy primary CRs.

The detection of EASs by employing ground detector arrays with large extensions (from 10$^{4}$ to 10$^{9}$ m$^{2}$) and long operation time (up to 20 years), extended the anisotropy studies up the highest energy region. The information about the individual shower was very limited but the very high counting rate, in particular for arrays located at high altitude, made this energy range attractive for anisotropy studies.  

The observation of anisotropy effects at a level of 10$^{-4}$ with an EAS array is a difficult job, because of how difficult is to control this kind of devices.
A wrong estimation of the exposure may affect the CR arrival rate distribution, even creating artifacts (i.e. fake excesses or deficit regions).
In fact, drifts in detector response and atmospheric effects on air shower development are quite hard to be modeled to sufficient accuracy\footnote{In fact, the temperature can modify the lateral extension of a shower trough the variation of the Moliere radius, and the pressure can influence the absorption of the electromagnetic component.}.
The envisaged solution is to use the data to estimate the detector exposure, but data contain either signal and background events, so that some distortions could be present in the results. 
The shape and the size of possible artifacts depend on the characteristic angle and time scale over which all the aspects of the data acquisition vary more than the effect to be observed.
Therefore, if an anisotropy of the order 10$^{-4}$ is looked for,
operating conditions must be kept (or made up) stable down to this level,
all across the field of view and during all the acquisition time.
In addition, as it is well known, the partial sky coverage carried out by a given experiment will bias the observations of structure on largest scales.
Finally, the observation of a possible small angular scale anisotropy region contained inside a larger one relies on the capability for suppressing the anisotropic structures at larger scales without, simultaneously, introducing effects of the analysis on smaller scales.

Both multi-directional and uni-directional detectors have been widely used to investigate different declination intervals defined by the center direction of viewing of the telescopes. 

Former experiments were not able to reconstruct the arrival direction of the primary, but the anisotropy induces periodicity in the response from a CR detector, having a fundamental frequency corresponding to the duration of a sidereal day. Physicists exploited the rotation of the Earth to build 1D distributions of the CR relative intensity.
Due to the poor statistics, any attempt to reconstruct the all-sky structure of the anisotropy was quite unrealistic and the only chance was to check if any strong statistical indication of a genuine CR intensity variation was there or not.

In the last decade, some experiments have been able to reconstruct the primary arrival direction with good precision on large amount of data, thus allowing to build two-dimensional (2D) maps of the anisotropy.
When 2D sky maps are realized, excesses and deficits observed in one dimension appear to be localized and to have their own extension and morphology. Any hypothesis about the correlation of the observed features with astrophysical sources can be confirmed or denied by the exact information on the direction. As 2D sky-maps for the CR anisotropy appeared only in the last few years, there is not any standard way to present data yet. Data are usually compared on the basis of the position and the extension of the observed features, as well as on their relative intensity. Recently, the IceCube experiments introduced a multi-scale analysis in spherical harmonics \cite{icecube11}.
\section{Experimental Results}
\label{sec:datareview}
\begin{figure}[!htbp]
\centering
\subfigure[]{\includegraphics[width=.9\textwidth]{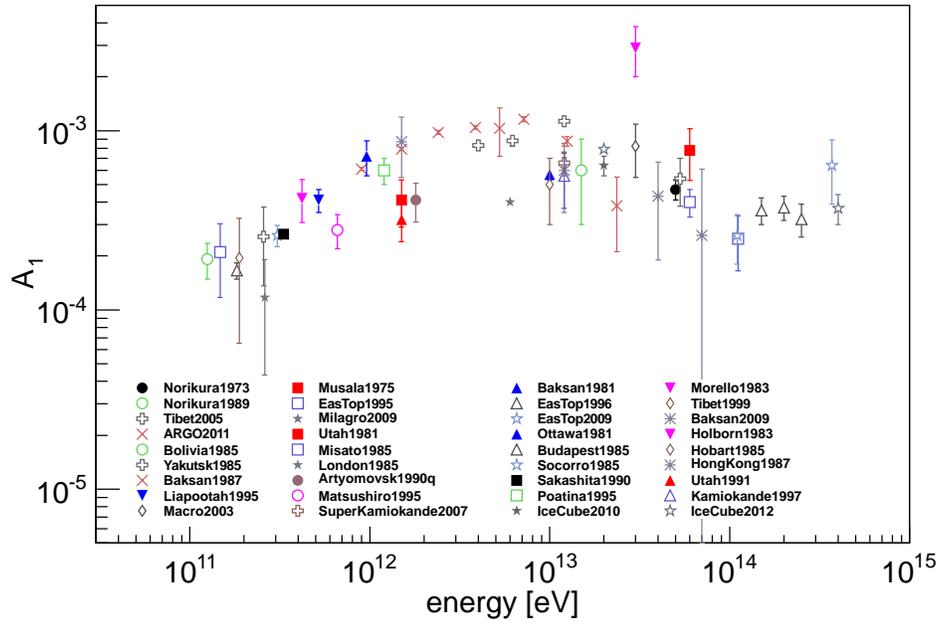}\label{fig:a}}
\subfigure[]{\includegraphics[width=.9\textwidth]{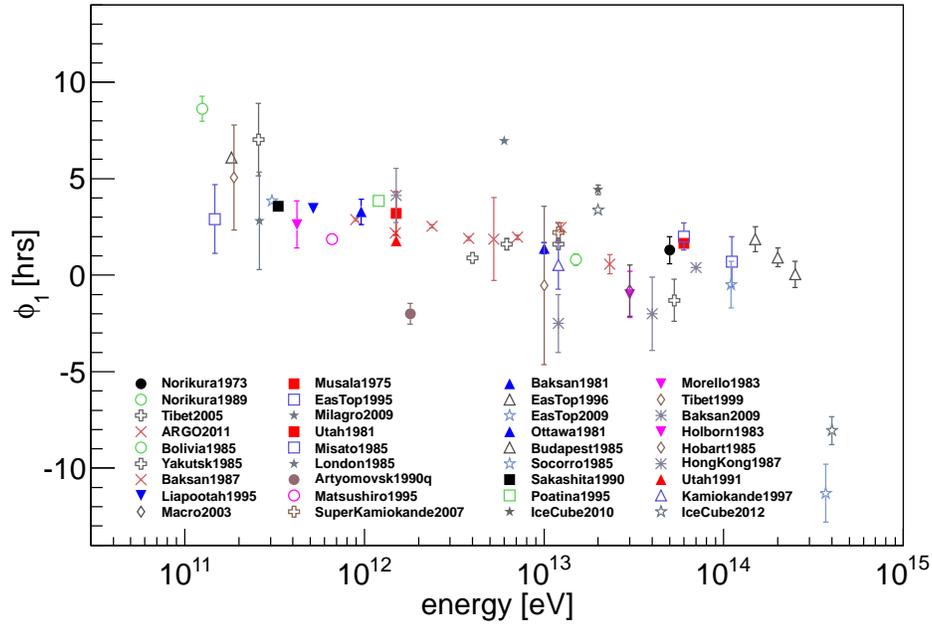}\label{fig:b}}
    \caption{First harmonics of the sidereal daily variations measured by underground muon detectors (\cite{swinson85,nagashima85a,ueno90,thambyaphillai83,munakata95,mori95,bercovich81,fenton95,cutler81,cutler91,lee87,bergamasco90,andreyev87,munakata97,superk07,macro03,icecube10,icecube12}) and EAS arrays (\cite{argo09,argo11,alexeenko81,munakata99,nagashima89,gombosi75a,amenomori05,milagro09,morello83,aglietta95,aglietta96,aglietta09}).
    The amplitude \subref{fig:a} and the phase \subref{fig:b} are plotted as a function of the primary CR energy.
For clarity, the points are not labeled with citation information.
 \label{fig:1amplitude}}
\end{figure}

In the figure \ref{fig:1amplitude} the amplitude and the phase of the first harmonic measured by different experiments starting since 1973 are shown as a function of the primary CR energy. 

A history of the experimental results is outlined in the next subsections.
%
%
%
\subsection{Observations pre-1975}
\begin{figure}[!ht]
  \centering
  \includegraphics[width=\textwidth]{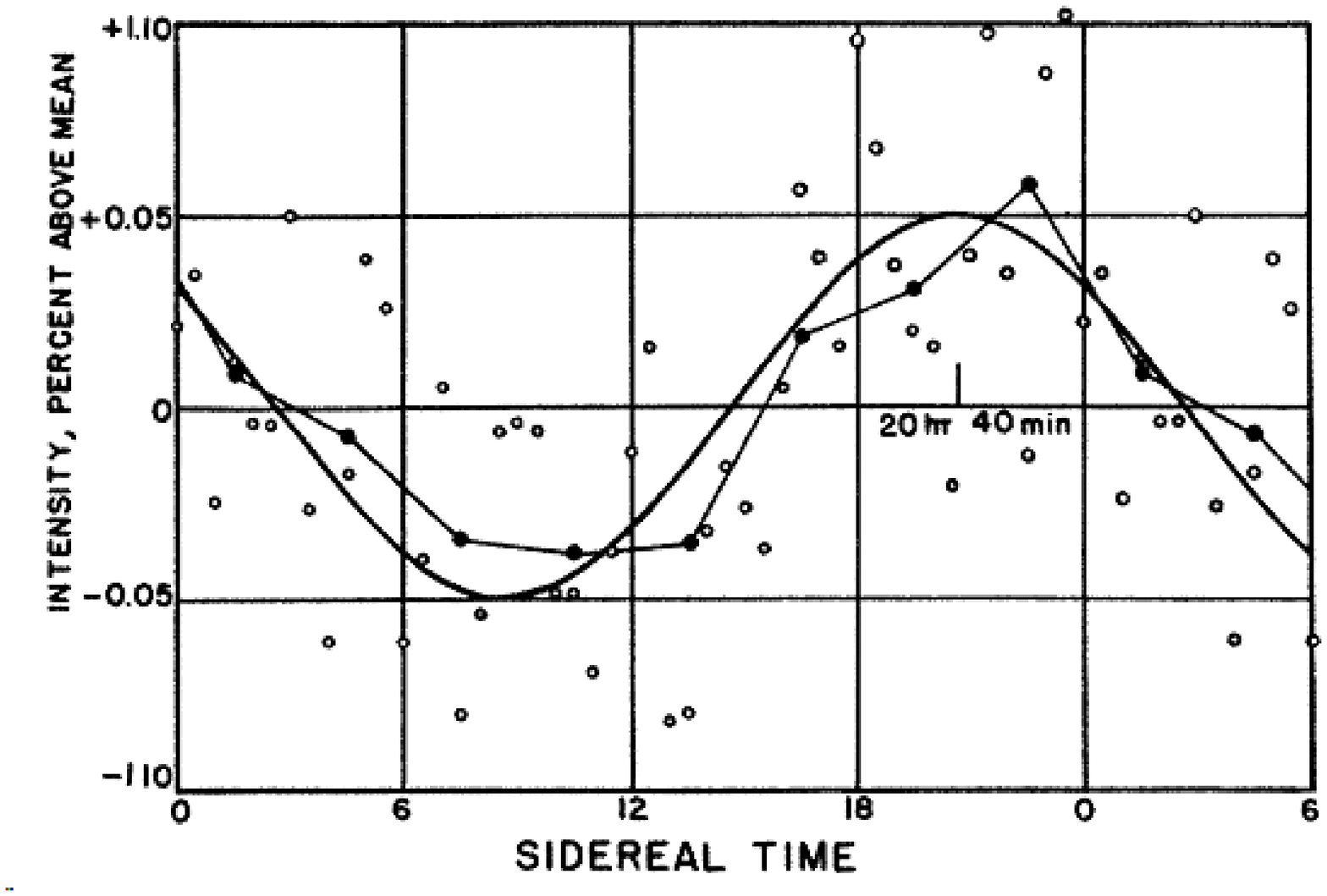}
  \caption{Percentage variation in intensity of the CRs with sidereal time. Data were collected by Hess and Steinmaurer in 1932. The open circles are half-our means; the solid circles are 3 hr means. The solid line represents the predicted Compton-Getting effect.
    \label{fig:hess}}
\end{figure}

The earliest evidence of periodicity in the CR intensity was reported in 1932 by Hess and Steinmaurer \cite{wollan39}, see the figure \ref{fig:hess}. Indeed, the question of a possible variation of the cosmic radiation flux with time arose soon after its discovery.
This observation, interpreted by Compton and Getting (CG) as an effect of the Earth motion relative to the sources of CRs \cite{comptongetting35}, stimulated many measurements to verify the CG model.
%
%

Although different ground-based experiments reported evidence of a diurnal variation of galactic CRs long before the 50s, solar modulation, the geomagnetic field and some atmospheric effects were considered a considerable obstacle for primary CRs of energy less than $10^{12}$ eV.
Indeed, this energy range, mainly investigated by underground muon telescopes, turned out to be very critical to get clear evidence on the anisotropy of CRs before they enter the solar cavity.

Variations due to the geomagnetic field or to the solar activity become negligible with increasing energy but, for a long time, the existence of the CR anisotropy in the multi-TeV energy range remained uncertain. In fact, the amplitude of the temporal variation of the EAS trigger rate (a few percent), mostly due to the atmospheric pressure and temperature effects, is much larger than the expected amplitude of the galactic CR anisotropy. The additional problem of the low event rate with increasing energy led to a number of contradictory results (see e.g., \cite{daudin53,escobar59,cachon62,gombosi75a}).

Only during 50s, when large detectors were operated underground or at the surface monitoring of the atmospheric variations, a clear evidence of the existence of the galactic anisotropy was obtained. 

In 1954 Farley and Storey \cite{farley54} published the results of the analysis of more than 10$^{5}$ EAS events recorded in Auckland, in the Southern hemisphere, with 3 counter trays at the corners of a 4 m-side triangle, in the period February 1951 - February 1952. They reported a significant observation of solar and sidereal diurnal variations of the CR intensity. 
By introducing the concept of anti-sidereal time an allowance were made, for the first time, for the seasonal amplitude modulation of the solar diurnal variation.
The comparison with two other experiments carried out in the northern hemisphere showed that the variations were in phase and the maxima coincident in local sidereal time. This gave important grounds for believing that the observed variation was a genuine sidereal effect and not a residual due to seasonal changes in the solar diurnal effect.

In 1957 two large wide angle muon telescopes were installed under 18 m of sandstone near Hobart \cite{fenton76} in the Southern hemisphere. The median primary energy was 1.8$\times$10$^{11}$ eV and the counting rate 7$\times$10$^4$ particles/hour.
Early in 1958 in the Northern hemisphere two almost similar muon telescopes were operated at a similar depth underground in Budapest \cite{somogyi76}.

\begin{figure}[!ht]
  \centering
  \includegraphics[width=\textwidth]{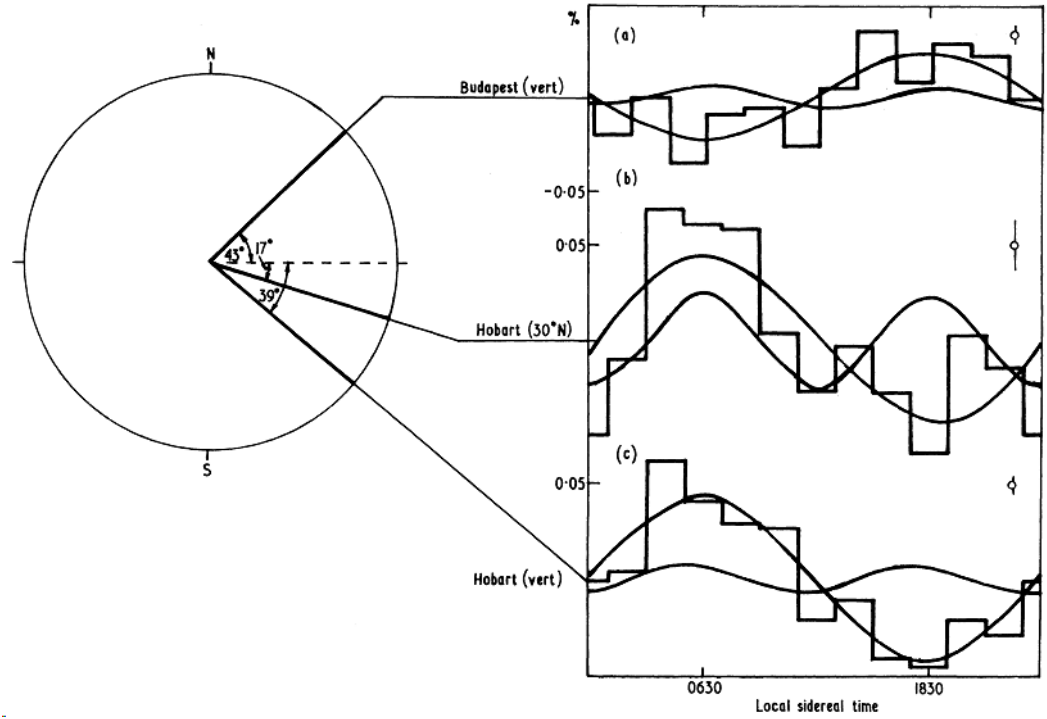}
  \caption{The sidereal daily variation of muon intensity observed underground between 1959 and 1962 in both hemispheres: (a) with vertical telescopes in Budapest, (b) with a telescope inclined 30$^{\circ}$ N in Hobart, (c) with vertical telescopes in Hobart. The first and second harmonics of best fit are shown. The latitude angles are those of the asymptotic directions of viewing. The error tails shown are the standard errors of the fitted amplitudes.} 
    \label{fig:hobartbuda}
\end{figure}

The combined analysis of the Hobart-Budapest data was very important to get evidence that the galactic anisotropy has higher order features besides the fundamental harmonic.
Observations of the sidereal daily variation revealed a latitude-dependent pattern in the first and second harmonics, as shown in the figure \ref{fig:hobartbuda}. A 12-hour difference between the phase of maxima in the two hemispheres was observed, as well as larger semi-diurnal maxima at the low latitudes. 
%
\begin{figure}[!ht]
  \centering
  \includegraphics[width=0.7\textwidth]{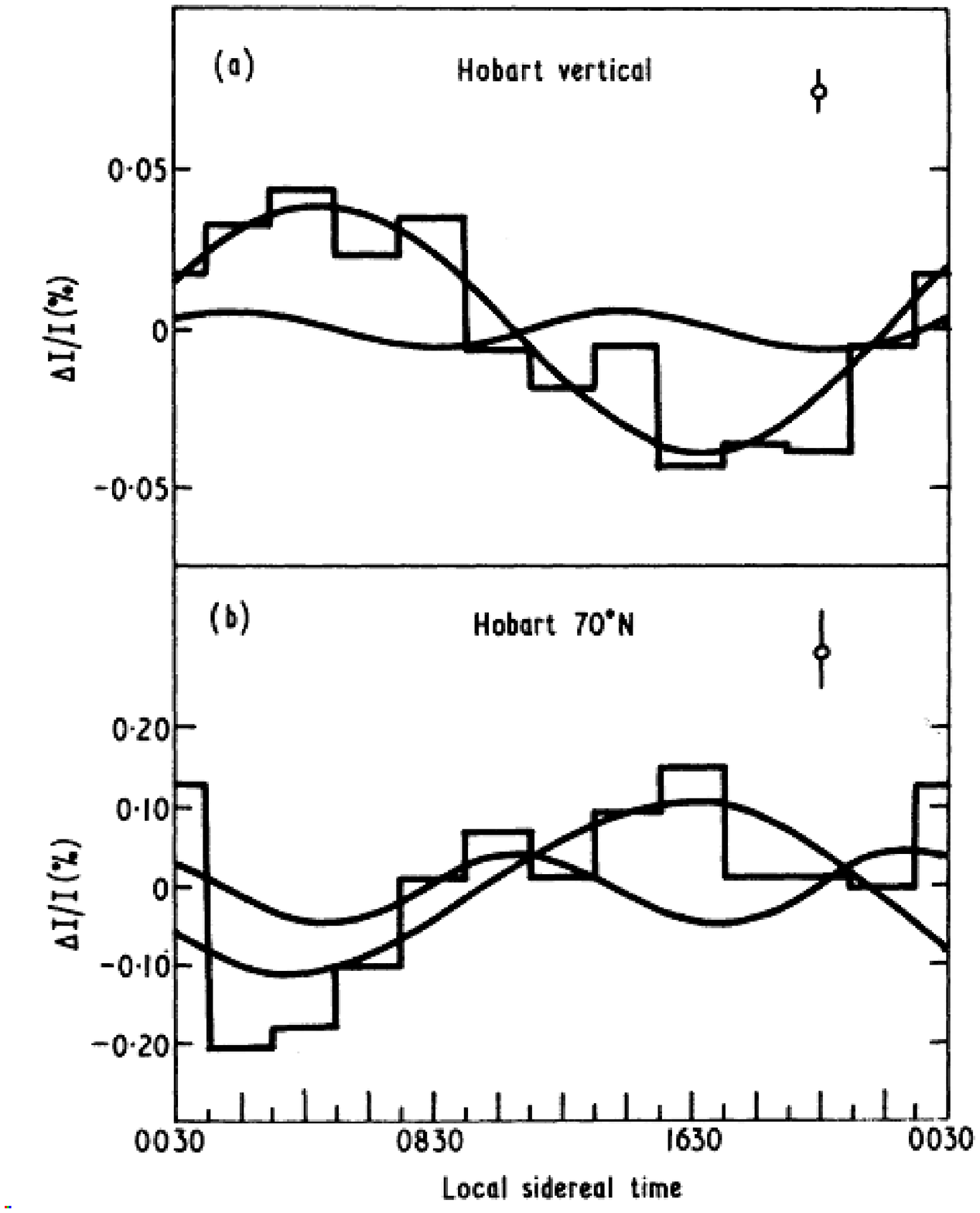}
\caption{The sidereal daily variation of muon intensity observed underground in Hobart (a) from vertical telescopes scanning an average southern latitude $\approx$ 39$^{\circ}$, (b) from a narrow angle telescope inclined 70$^{\circ}$ N of zenith, scanning an average northern latitude $\approx$ 20$^{\circ}$. The first and second harmonics of best fit are shown. The error tails shown are the standard errors of fitted amplitudes.}
  \label{fig:hobart}
\end{figure}
%

These results suggested the existence of a global galactic anisotropy consisting of a collinear types of bidirectional anisotropy and a smaller unidirectional anisotropy in line with it. The particular bidirectional type was indicated by the diurnal maxima in phase with one of the semi-diurnal maxima and the remarkable dependence on the latitude of the intensity of the semi-diurnal modulation. The fact that the diurnal amplitude in Hobart was considerably larger than in Budapest suggested the presence of a unidirectional diurnal variation in phase with the bidirectional component in Hobart and therefore 12 hours out of phase with that in Budapest \cite{jacklyn65,jacklyn86}.

However, the 12-hour phase difference between the two hemispheres naturally suggests the possibility that the result is not due to a genuine sidereal anisotropy but to a spurious sidereal effect arising from the seasonal modulation of the solar daily variation of CR intensity.
To test the difference between the two experiments a narrow angle telescope was installed in Hobart pointing north at a sufficiently high angle (70$^{\circ}$) to the zenith so as to view into the northern hemisphere. The telescope observed a sidereal daily variation not in phase with the vertical telescope in Hobart, but with the Budapest phase, see the figure \ref{fig:hobart}, thus excluding the presence of a spurious component \cite{jacklyn65}.

A very large narrow angle air Cerenkov telescopes operated in Nagoya confirmed these results recording the variation of CR intensity at a median energy of about 2$\times$10$^{11}$ eV in the period 1968-1971 \cite{sekido71}.
Further evidences came from a couple of narrow angle telescopes operated in Mawson in 1968 \cite{jacklyn69,jacklyn71}.

From 1965 onwards a great increase took place in the number and size of underground experiments. 
A highly significant development in the early 1970s was the start of observations of small air showers at a median energy around 10$^{13}$ eV at Mt. Norikura in Japan \cite{sakakibara73}.

In fact, the observations made in the energy range 10$^{13}$ - 10$^{15}$ eV until that moment did not give concrete evidence for the existence of the sidereal anisotropy due to several reasons. First, the counting rates attained was not sufficient to obtain the necessary accuracy of the order of 10$^{-4}$. Moreover, the experiments were carried out with no consideration of the stability of the apparatus, thus with dispersion of actual data far greater than the statistical one.

In the Mt. Norikura experiment these problems were worked around because the apparatus was a multi-directional telescope housed in an air-conditioned laboratory with daily temperature variation less than $\sim$0.1$^{\circ}$C. In this way, th effect of the temperature on the detector was controlled to be less than $\sim$0.01\%. The high counting rate ($\sim$45,000/hour) was attained due to the large detector area and the high altitude location (2770 m a.s.l.).
In addition, to reduce the atmospheric effects, especially that of external temperature variations, the observation of directional air showers (East and West component) was carried out utilizing the absorption effect of air showers by a 10 cm lead wall \cite{sakakibara75}.
The omni-directional Norikura array measured a sidereal anisotropy consistent with the Hobart-Budapest results, thus making the scientific community confident that no significant spurious component had affected the experimental results. The importance of this measurement was due to its being the first directional-measurement of anisotropy ever.

The Mt. Norikura observations were confirmed by other EAS arrays from Peak Musala in Bulgaria and from Baksan in Georgia, in the energy range 10$^{13}$ - 10$^{14}$ eV \cite{sakakibara84}. 
In 1970s another important experiment, made of three vertical telescopes for muon detection, started operations in Poatina, Tasmania, at a median energy of about 5$\times$10$^{11}$ eV.
Other important results came from a group of large muon-directional plastic scintillator telescopes located at ground in Nagoya and underground in Misato and Sakashita \cite{ueno84,ueno85}.

Nevertheless, in spite of the progress in detection techniques, the reported amplitudes were so small in relation to factors affecting observations (geomagnetic and helio-magnetic fields, solar modulations, temperature and pressure variations, low counting rates) that their accuracy as a measure of the CR anisotropy remained in doubt for many years and the results not conclusive yet. 
\subsection{Observations post-1975}
The Munich International Cosmic Ray Conference (ICRC) in 1975 marked a change in viewpoint regarding the CR arrival direction anisotropy \cite{fenton75a,linsley83}. In fact, in three papers by Gombosi et al. \cite{gombosi75b}, Nagashima et al. \cite{naga75} and Fenton and Fenton \cite{fenton75b}, the observation of a sidereal anisotropy seen with approximately the same amplitude ($\sim$ 0.07\%) and phase ($\sim$ 00 hr RA) at approximately the same latitude in both hemispheres was reported in the energy interval 10$^{12}$ - 10$^{14}$ eV/nucleon. 
In addition, in this conference there was a clear experimental evidence that solar effects overwhelm the results of underground telescopes detecting muons produced by primary CRs with energy less than $\leq$10$^{12}$ eV/nucleon. Such effects prevented any clear determination of the anisotropy from being made in this energy range \cite{fenton75a}.

In 1977 Linsley and Watson reviewed the data on CR anisotropy pre-1965 in the energy range 10$^{14}$ - 3$\cdot$10$^{17}$ eV concluding that the combined results cannot be explained by random fluctuations or selection effects \cite{linsley77}. 

In the following years other measurements carried out at Musala peak \cite{gombosi77}, Mt Norikura \cite{sakakibara76,sakakibara79} and in Baksan \cite{alexeenko81} provided additional confirmation of the existence of a sidereal CR anisotropy. 
In addition these results from EAS-arrays in the energy range 10$^{13}$ - 10$^{14}$ eV were in good agreement with observations from experiments with underground muon telescopes around 10$^{12}$ eV \cite{fenton76,davies79,bergeson79,bercovitch81}.

In 1983 Linsley concluded that the results obtained by a harmonic analysis in sidereal time ``should be accepted without any reservations about spurious effects of a statistical, instrumental or atmospheric nature. (...) 
The success of the small air shower experiments shows that the residual effect of instrumental instabilities can be held to 1 part in 10$^4$'' \cite{linsley83}.

In the mid 80s, new analysis techniques for underground muon observations were developed, to eliminate the spurious sidereal variation from the apparent variation. Those methods were based on the observed anti-sidereal diurnal variation \cite{nagashima85a}, which was demonstrated to be capable to eliminate effects on the low energy CRs due to the reversal of the Sun's polar magnetic field \cite{nagashima85b}. 

The same methods were soon applied to the observation of EASs conducted since 1970 with an air shower detector at Mount Norikura in Japan (median energy $\approx$ 1.5$\times$10$^{13}$ eV).
A significant sidereal diurnal variation at higher energy was observed, together with the strong evidence of semi-diurnal and tri-diurnal variations \cite{nagashima89}. As semi-diurnal and tri-diurnal modulations correspond to 180$^{\circ}$ and 120$^{\circ}$ periodicity respectively, excesses as wide as 60$^{\circ}$ were found, showing some hints of sub-structures and medium scale features already at that time.

A continuous observation of the 24 hr profile of the CR rate in sidereal time over 12 years revealed that the sidereal diurnal variation of 10 TeV CR intensity exhibits a deficit with a minimum around 12 hr \cite{nagashima89}. A similar anisotropy was reported in those years by other experiments in the same energy region \cite{alexeenko85,bergamasco90}, and also down to $6\times10^{10}$ eV, by underground muon detectors, with the same form and almost the same phase.

Until this moment, the observation of the galactic anisotropy was concerned with two types, unidirectional and bidirectional \cite{jacklyn86}. They may occur independently of each other or as a component of a more complex global anisotropy. The unidirectional anisotropy can be produced by the motion of the solar system relative to a frame of reference in which the CR system can be regarded as isotropic (the CG effect). The diurnal variation is symmetric, in the sense that it is the same at the same latitude in both hemispheres.
On the contrary, a bidirectional diurnal variation is asymmetric with respect to the hemisphere, since times of maximum will differ by 12 hr between northern and southern latitudes of observation. Consequently, if the total observed diurnal variation consists of superimposed unidirectional and bidirectional components they can be separately identified by taking the sum and difference of observations, made under the same detecting conditions, at the equal latitudes in both hemispheres. The unidirectional component is separated out in the sum of observations and the bidirectional one in the difference.

In 1998 Nagashima, Fujimoto, and Jacklyn (NFJ hereafter) reported the first comprehensive observation of a large angular scale anisotropy in the sub-TeV CRs arrival direction by combining data from different experiments in the northern and southern hemispheres \cite{nagashima98}. 
They analyzed data of the muon telescopes underground in Sakashita and Hobart (median energies $\approx$ 331-387 GeV and $\approx$ 184 GeV, respectively), and at the Nagoya ground station ($\approx$ 60-66 GeV). Data recorded with the small air shower array at Mt. Norikura were also used.

\begin{figure}[!ht]
  \centering
  \includegraphics[width=\textwidth]{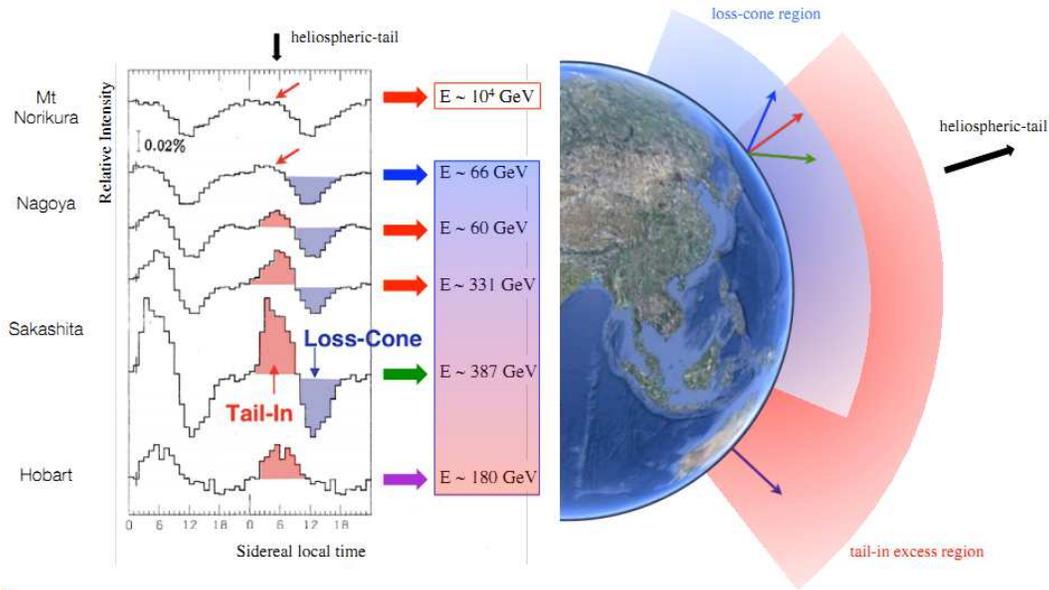}
  \caption{CR sidereal daily variation from different muon telescopes or air shower detectors: from top to bottom, Mount Norikura air shower array (shown for reference at high energy and with the amplitude multiplied by 1/4 to account for the larger anisotropy amplitude at high energy), Nagoya station (looking 30$^{\circ}$ northward, and looking vertically upward), Sakashita station (looking vertically upward and 41$^{\circ}$ southward), and Hobart station (looking vertically upward). The CR median energy of different detectors is shown. In order to clearly show the peaks and valleys, the 24 hr variation is repeatedly shown in a 2-day time interval. The error expresses the dispersion of the hourly relative intensity (courtesy of P. Desiati).
    \label{fig:nfjobs}}
\end{figure}

The results obtained by the NFJ analysis are shown in the figure \ref{fig:nfjobs}.
In addition to a \emph{di-polar anisotropy} (the CR deficit around $\alpha$ = 0 hr $\delta$ = -20$^{\circ}$, with intensity independent of the energy), which they named ``galactic'' anisotropy, NFJ found a \emph{new anisotropy component} causing an excess of intensity with a maximum around 6 hr in the sub-TeV region.
%
%
 
This directional excess flux, confined in a narrow cone with a half opening angle of $\sim$ 68$^{\circ}$, seemed to coincide with the expected helio-magneto-tail direction ($\alpha$ = 6.0 hr $\delta$ = -29.2$^{\circ}$), opposite to the proper motion of the solar system and it was supposed to be of solar origin and was named \emph{``tail-in''} by NFJ.

The amplitude of these anisotropy components increases with energy up to $\approx$ 1 TeV. After this energy, the new contribution was observed to be less and less important, even if disappearing at $\sim$10 TeV (Mt Norikura).

NFJ suggested that such a decrease is consistent with a signal related to the acceleration of particles arriving from the helio-tail direction, i.e., $\sim$ 6 hr in sidereal time.

It was an indirect confirmation that higher energy particles are an ideal tool to investigate the galactic anisotropy, being insensitive to Sun-related effects.
%
\begin{figure}[!ht]
  \centering
  \includegraphics[width=0.8\textwidth]{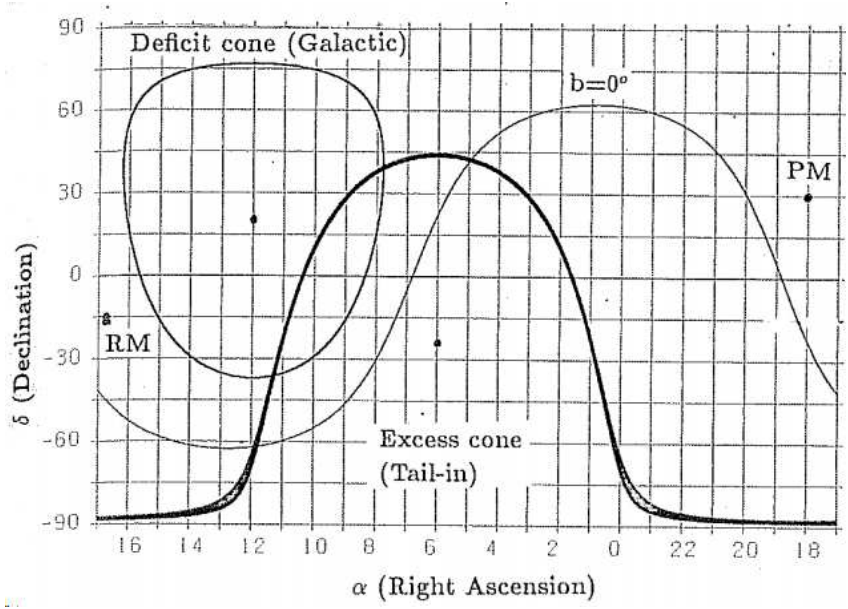}
  \caption{The distribution of the tail-in and loss-cone (galactic) anisotropies on the equatorial coordinate grid. The excess cone of the tail-in anisotropy, within which all the directional excess flux is confined, is shown bounded by the thick line including the south pole. A thinner line bounds the deficit cone of the galactic anisotropy, in which all the directional flux is confined. PM shows the direction of the proper motion of the solar system. RM shows the direction of the relative motion of the system to the neutral gas. The galactic equator is represented by the thin line labelled by $b=0^{\circ}$.
    \label{fig:nfjmodel}}
 \end{figure}
%

In the earliest ``map'' of the large scale anisotropy, figure \ref{fig:nfjmodel}, referred to as the \emph{NFJ model}, the excess and deficit cones were obtained by interpolating between one dimensional anisotropy measurements made in several different declination strips. 

the result was qualitative in nature because data were from different detector types (shallow underground muon telescopes and surface EAS array) with large spread in energy sensitivity and different systematic uncertainties.

The tail-in excess was recognized to be not symmetric along the R.A. direction and this result confirmed the discovery of the semi- and tri-diurnal variations ten years before. Some long-period measurements from neutron's monitors did not observe any decrease of the tail-in excess amplitude during the reversal of the solar magnetic field from the negative to the positive polarity state (1989). This result strengthened the hypothesis of the ``helio-spheric'' origin of the phenomenon, against the ``galactic'' one \cite{Karapetyan_neutron_anis_2010}.

The positive pole of the di-polar anisotropy slightly overlaps with the tail-in zone, what makes the deficit di-polar region more evident alone. This region is often referred as \emph{``loss-cone''}.

The discovery of the tail-in gave an important piece for the solution of the problem of the observed phase shift in the sidereal diurnal variation, from 6 to 0 hours with the increase of the energy. 
In fact, NFJ explained the phenomenology by distinct contributions from two anisotropies. 
The amplitude of the tail-in anisotropy, with a maximum at $\sim$ 6 hr in sidereal time, decreases with increasing primary energy above $\sim$ 1 TeV, while the Galactic anisotropy, with a minimum at $\sim$ 12 hr, remains constant.
Furthermore, NFJ suggested that the phase of the composite first harmonic vector turns counterclockwise from $\sim$ 3 to $\sim$ 0 hr with increasing energy (the composite amplitude is also expected to decrease by $\sim$ 30\% when the amplitudes of two anisotropy components are equal at $\leq$ 1 TeV). 

Finally, it was clearly demonstrated that the observed sidereal variation was not compatible with the CG effect expected from the motion of the solar system in the interstellar space. The authors drew the conclusion that the solar system drags with it in its motion the surrounding interstellar magnetic field within which the CRs with low energy are isotropically confined.

The work \cite{nagashima98} was so important to make the ``tail-in'' and ``loss-cone'' names used by the scientific community, although there is no agreement yet about their origin.

A detailed analysis of the sidereal diurnal modulations observed in a total of 48 directional channels of the underground muon detectors monitoring both the northern and southern sky indicated that the maximum phase of the new anisotropy component, which is $\sim$6 hr in the northern hemisphere, shifts toward earlier times as the declination of the incident CRs moves southward to the equator \cite{hall98,hall99}.
%
\begin{figure}[!ht]
  \centering
  \subfigure[Thick black lines: counting rate curves in solar (a), sidereal (b), and anti-sidereal (c) time at 1.1$\times$ 10$^{14}$ eV. The statistical uncertainty for each bin is given in the first one. The curves resulting from the first harmonic analysis are also shown (light black lines); for the sidereal time curve, the combination of the first and second harmonics (dotted black line) is additionally superimposed.]{
    \includegraphics[width=0.42\textwidth]{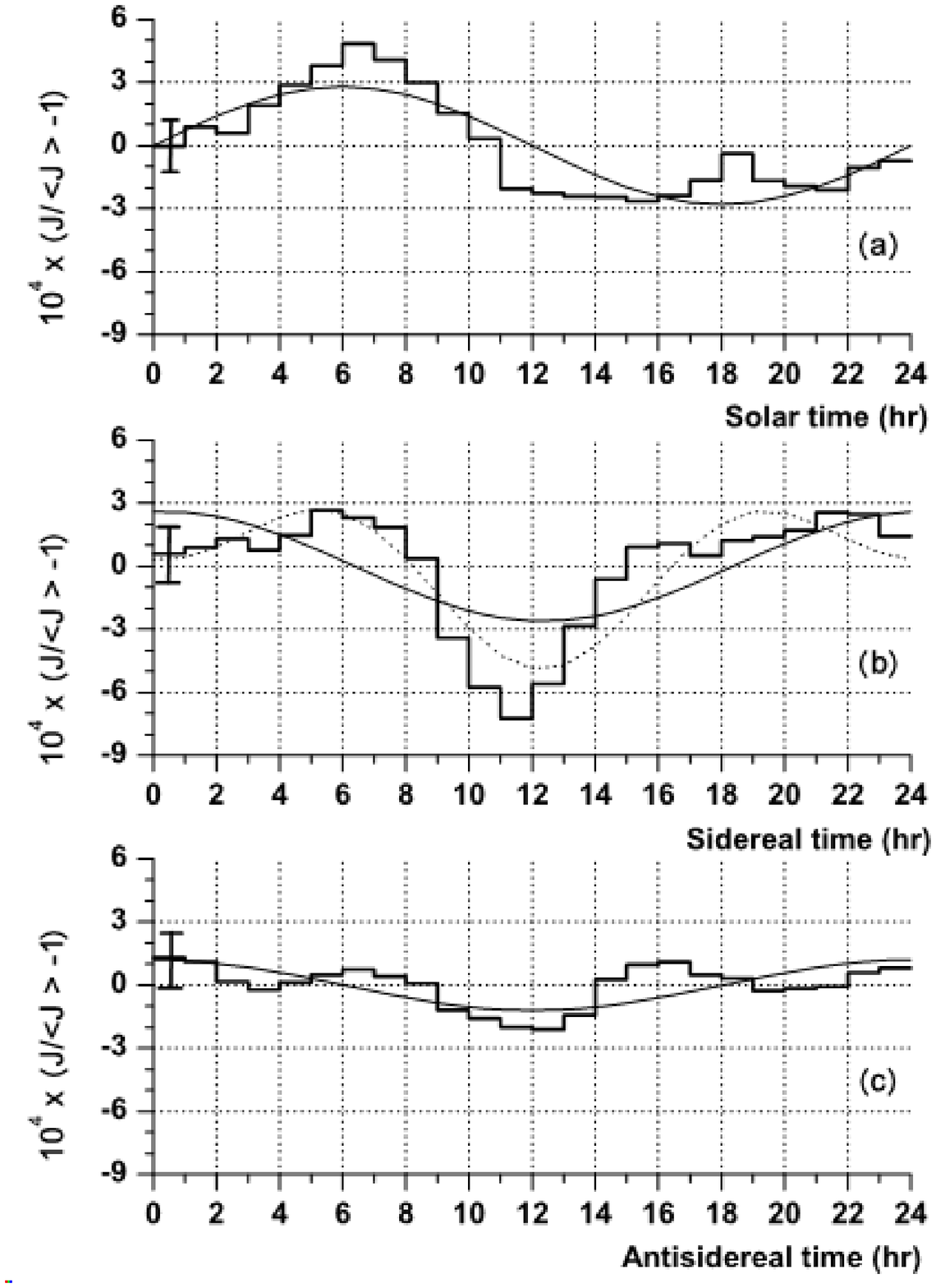}
    \label{fig:eastop}}
\hspace{0.1\textwidth}
  \subfigure[Thick black lines: counting rate curves in solar (a), sidereal (b), and anti-sidereal (c) time at 3.7$\times$10$^{14}$ eV. The curves resulting from the first harmonic analysis are also shown (light black lines).]{
    \includegraphics[width=0.42\textwidth]{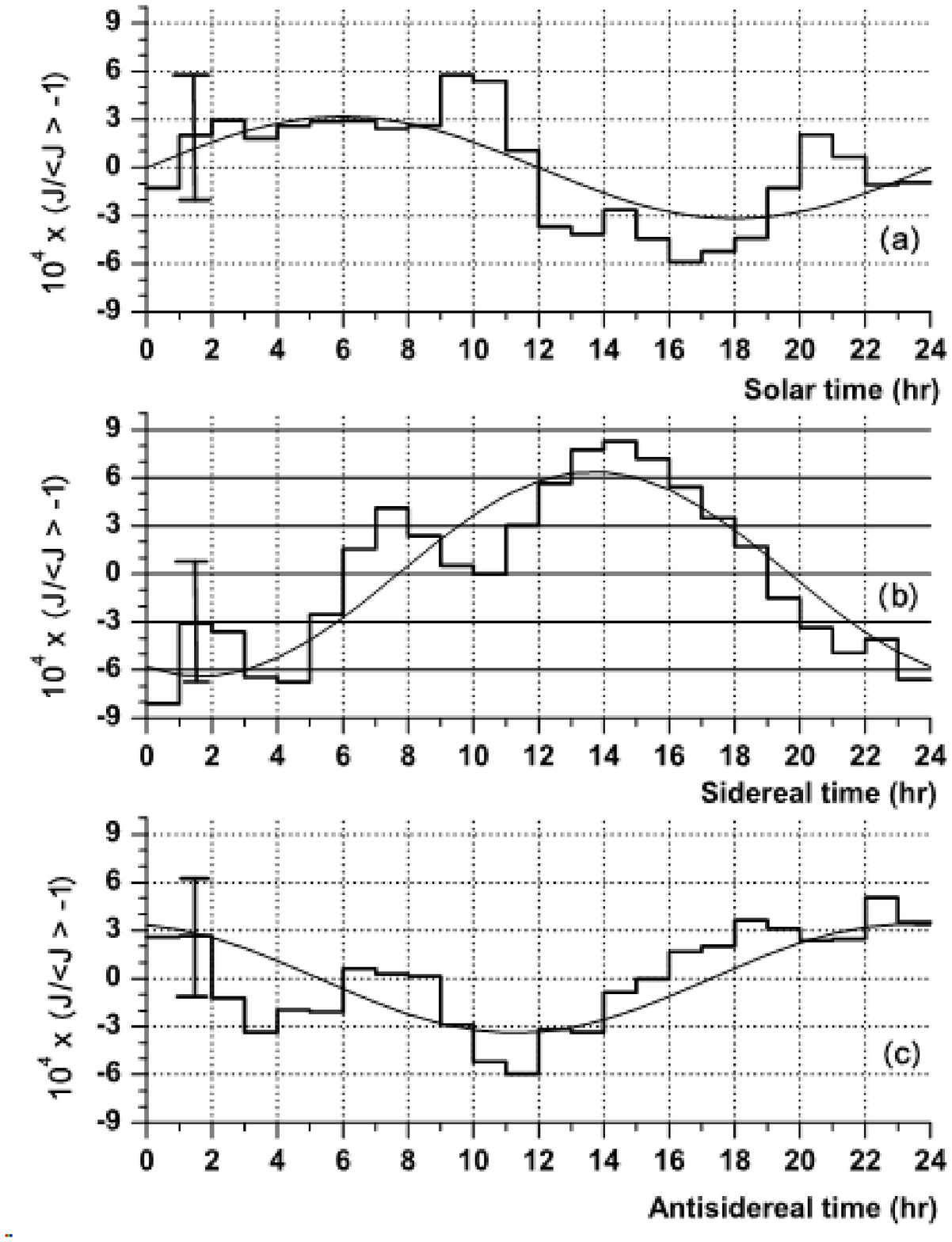}
    \label{fig_eastop400t} }
\end{figure}
%

In 1996 the EAS-TOP collaboration reported a detailed analysis of high energy CR anisotropy at $\approx10^{14}$ eV. Data were analyzed in solar, sidereal and anti-sidereal time using the East-West method, see the figure \ref{fig:eastop}. A signal compatible with the CG effect is clearly observed in solar time, for the first time at energy $>10^{14}$ eV, whereas no significant structure was observed in anti-sidereal time \cite{aglietta96}.
In 2009 the EAS-TOP collaboration reported the first evidence ever of anisotropy around 400 TeV \cite{aglietta09}.
Hints of increasing amplitude and change of phase above 100 TeV resulted from the analysis of the full data set, as shown in the figure \ref{fig_eastop400t}.

In the same period the underground MACRO experiment, measuring $>$10 TeV muons, found evidence for modulations ($<10^{-3}$) in solar and sidereal diurnal periods, at the limit of the detector statistics. The solar diurnal modulation was attributed to the daily atmospheric temperature variations at 20 km, the altitude of primary CR interaction with the atmosphere \cite{macro03}.

From the measurements done by ``old'' experiments (see \cite{hillas84,fichtel86} for a review) it was claimed that the amplitude of the first harmonic increases with increasing energy, starting from 10$^{15}$ eV. The phase too seemed to be energy dependent, with changes correlated with the Galactic structure. However, the statistical uncertainty was too large to drew any conclusions. In fact, the anisotropy apparent amplitude naturally increases as the number of events decreases (i.e., as the energy increases), as a simple consequence of (lack of) statistics: $\delta$ = (I$_{max}$ - I$_{min}$)/(I$_{max}$ + I$_{min}$) $\propto$ N$^{0.5}$/N = 1/N$^{0.5}$ \cite{ghia07}.
%
\begin{figure}[h]
  \begin{center}
    \mbox{\epsfig{file=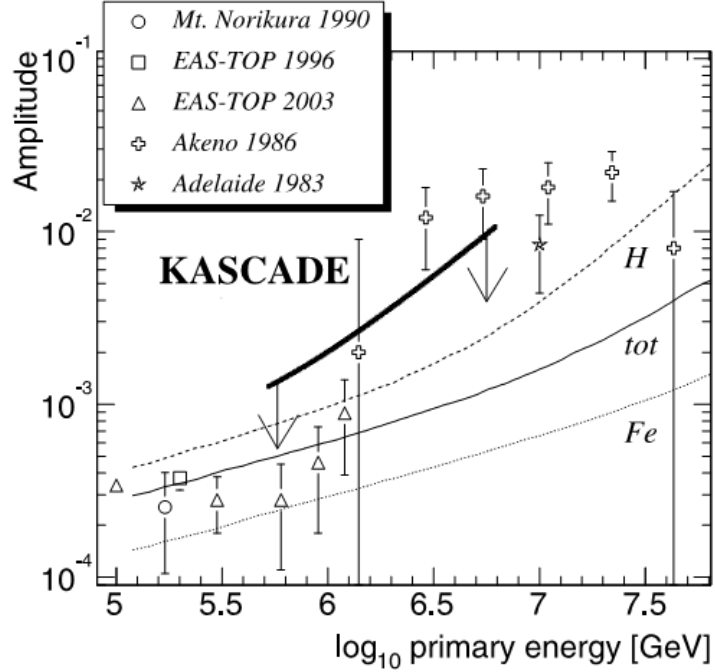,width=0.7\textwidth}}
  \vspace{-0.5pc}
    \caption{KASCADE upper limits (95\%) of Rayleigh amplitudes vs.
primary energy (thick line) compared to other results \cite{nagashima90,aglietta96,aglietta03,akeno86,adelaide83}. Model predictions from \cite{candia03} for the total anisotropy and for the anisotropies of the proton and iron components are also shown (thin lines).
 \label{fig:kascade}}
  \end{center}
\end{figure}
%

In 2004 the KASCADE experiment presented an analysis of the CR anisotropy in the PeV energy range. 
No hints of anisotropy in the R.A. distributions were reported and upper limits for the Rayleigh amplitude between 10$^{-3}$ at a primary energy of 0.7 PeV and 10$^{-2}$ at 6 PeV were set, see the figure \ref{fig:kascade}  \cite{kascade04}.

The EAS-TOP collaboration in 2003 published a measurement of the first harmonic parameters up to 1.2 PeV \cite{aglietta03}. The amplitudes are shown in the figure \ref{fig:kascade}. Since the signal above 300 TeV was not statistically significant, they set upper limits ranging between 4$\cdot$10$^{-4}$ at 0.3 PeV and 2$\cdot$10$^{-3}$ at 1.2 PeV.
We recall what already mentioned in this section, i.e. in 2009 the EAS-TOP collaboration published a new analysis providing the amplitude and phase of the first harmonic at 370 TeV.

The EAS-TOP and KASCADE experiments did not confirm the increase of anisotropy with energy (from 10$^{15}$ eV upward) claimed by old EAS arrays, though limited statistics still prevents to draw any firm conclusion.
\subsection{Recent Observations: 2D Sky Maps}
Based on a harmonic analysis, these observations are consistent with the large-scale diffusive propagation of CRs in the Galaxy \cite{shibata04}, but there was no consensus for a production mechanism in the local interstellar region surrounding the heliosphere that would reproduce the sidereal diurnal variation. 

In fact, although a general view of the situation could be obtained by merging data from different experiments (e.g., see the figure \ref{fig:nfjmodel}) the lack of homogeneous (i.e., one-experiment) information on the dependence of the anisotropy on declination impeded a complete experimental formulation of the problem \cite{amenomori05}. 

On the other hand, the 24 hr profile and its declination dependence can be measured \emph{together} only with direction-sensitive devices.

In the last decade large area detectors with correspondingly large statistics and good pointing accuracy came on line, each one able to make a two-dimensional (2D) representation of the CR arrival direction distribution, thus allowing detailed morphological studies of the anisotropy either along R.A. and declination directions.
%
\begin{figure}[!ht]
  \begin{center}
    \mbox{\epsfig{file=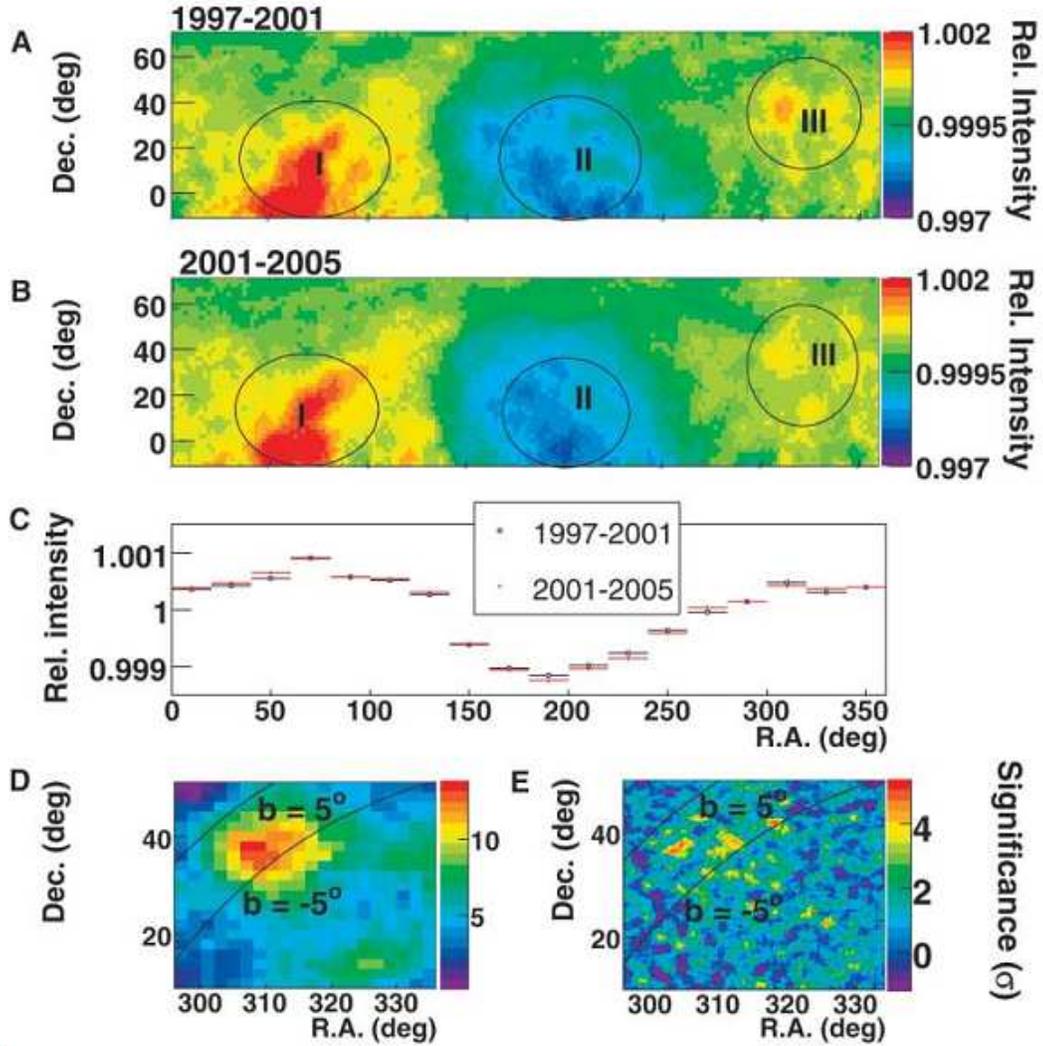,width=\textwidth}}
  \vspace{-0.5pc}
    \caption{Equatorial CR intensity map for Tibet AS$\gamma$ data taken from (A) 1997 to 2001 and (B) 2001 to 2005. The vertical color bin width is 2.5$\times$ 10$^{-4}$ for the relative intensity in both (A) and (B).
The circled regions labeled by I, II, and III are the tail-in component, the loss-cone component, and the newly found anisotropy component around the Cygnus region, respectively. (C) The 1D projection of the 2D maps in R.A. for comparison. (D) and (E) show significance maps of the Cygnus region for data from 1997 to 2005. Two thin curves in (D) and (E) stand for the
Galactic parallel b$\pm$5$^{\circ}$. Small-scale anisotropies (E) superposed onto the large-scale anisotropy hint at the extended gamma-ray emission.
 \label{fig:tibet06}}
  \end{center}
\end{figure}
%

In 2006 the Tibet AS$\gamma$ experiment, located at Yangbajing (4300 m a.s.l.), published the first 2D high-precision measurement of the CR anisotropy in the Northern hemisphere in the energy range from few to several hundred TeV \cite{amenomori06}. in the figure \ref{fig:tibet06} the CR intensity maps observed by Tibet AS$\gamma$ are shown. The 1D projection of the 2D maps in R.A. is also shown. This measurement revealed finer details of the known anisotropy and established a new component of the galactic anisotropy around the Cygnus region (region III of panel D). 

In solar time the observed di-polar anisotropy was in fair agreement with the expected CG effect due to the Earth's orbital motion around the Sun.

For CR energy higher than a few hundred TeV, all the components of the anisotropy fade out, showing a co-rotation of galactic CRs with the local Galactic magnetic environment.
%
\begin{figure}[!ht]
  \begin{center}
    \mbox{\epsfig{file=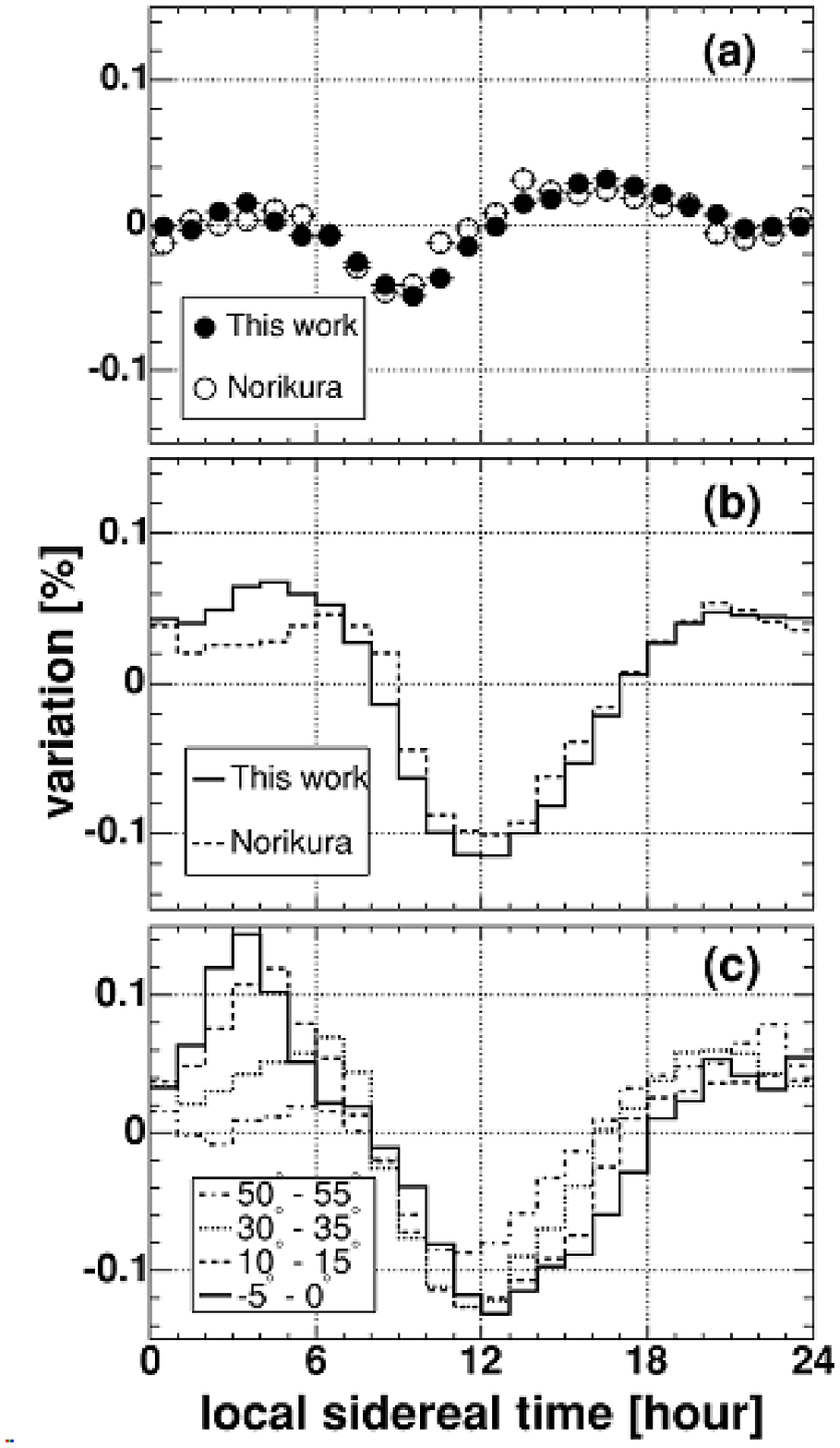,width=0.67\textwidth}}
  \vspace{-0.5pc}
    \caption{Sidereal daily variations averaged over all primary energies. Panels (a) and (b) show, respectively, the differential variation and the physical variation averaged over all declinations, while panel (c) shows the physical variation in four declination bands. The open circles in (a) and the dashed histograms in (b) display the variations reported from the Norikura experiment (for details see \cite{nagashima89}).
 \label{fig:tibet05}}
  \end{center}
\end{figure}
%

The Tibet AS$\gamma$ collaboration carried out the first measurement of the energy and declination dependences of the R.A. profiles in the multi-TeV region with a single EAS array \cite{amenomori05}.
They found that the first harmonic amplitude is remarkably energy-independent in the range 4 - 53 TeV, contrary to the suggestions of the NFJ model. 

The panels (a) and (b) of figure \ref{fig:tibet05} show the full 24 hr profile of the sidereal diurnal variation averaged over all declinations and primary energies, together with the results of the Mt Norikura EAS array, which was not capable of resolving the declination of the incident direction. 
The Tibet AS$\gamma$ results are in good agreement with the Norikura profile, if the average over all declinations is considered. 

The importance of this new result became even clearer when looking at the panel (c), where the sidereal profiles in each declination band separately are shown. 
The phase of maximum changes from $\sim$ 7 to $\sim$ 4 hr in sidereal time, and the amplitude of the excess increases as the viewing direction moves from the northern hemisphere to the equatorial region. This is consistent with the anisotropy component firstly found in the sub-TeV region by underground muon detectors and referred to as tail-in anisotropy.
The Tibet AS$\gamma$ experiment clearly showed that the tail-in anisotropy continues to exist in the multi-TeV region, what have deep implications for all models of CR acceleration and propagation providing predictions on the anisotropy.
%
\begin{figure}[!ht]
  \begin{center}
    \mbox{\epsfig{file=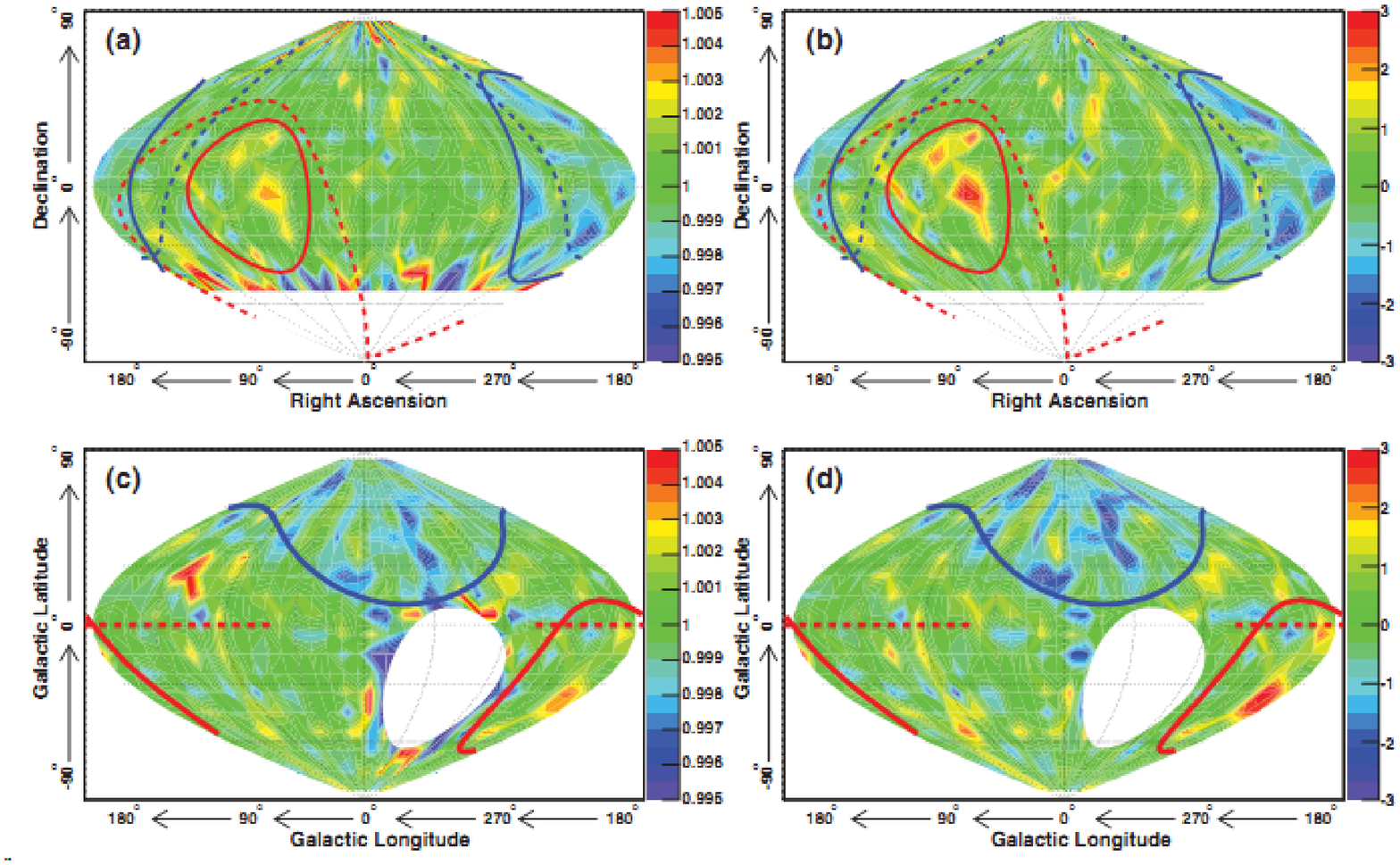,width=0.8\textwidth}}
  \vspace{-0.5pc}
    \caption{Sky map of the anisotropy in equatorial coordinates. The sky is divided into 10$^{\circ}\times$ 10$^{\circ}$ cells. Declinations less than -53.58$^{\circ}$ (white region) always lie below the horizon and are thus invisible to the detector. In (a), each cell shows the fractional variation from the isotropic flux, while in (b) it shows the standard deviation of this variation. The solid red and blue curves show the excess and deficit cones obtained using a clustering algorithm applied to the data. The dashed curves in (a,b) show excess and deficit cones from the NFJ model. (c) and (d) show the maps in (a) and (b) transformed to the galactic coordinates. The solid red and blue curves are the same cones as described above. The dashed red horizontal line indicates the direction of the Orion arm. The white patch indicates the below-horizon region.
 \label{fig:sk1}}
  \end{center}
\end{figure}
%

Though with less sensitivity, a 2D map of the CR anisotropy was reported also by SuperKamiokande-I (SK-I) in 2007 \cite{superk07}. The experiment, realized  with an underground imaging water Cerenkov detector, published the map shown in the figure \ref{fig:sk1}, representing high energy muons ($>$ 10 TeV) coming from different sky regions. 
The observation confirmed the existence of two large scale anisotropy regions, an excess and a deficit region, in agreement with the NFJ model and with the Tibet AS$\gamma$ measurement.
%
\begin{figure}[!ht]
  \begin{center}
    \mbox{\epsfig{file=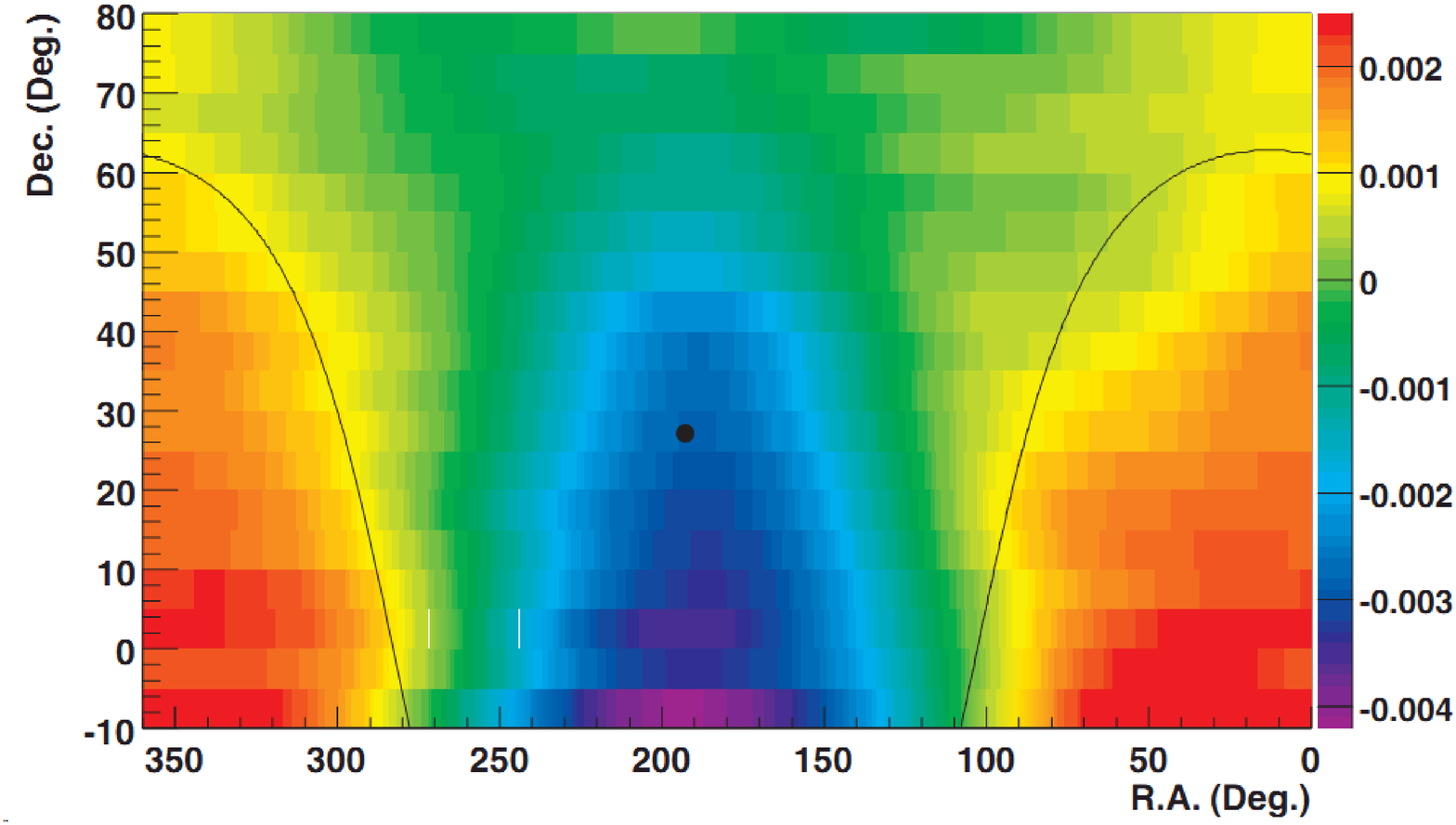,width=0.8\textwidth}}
  \vspace{-0.5pc}
    \caption{Result of a harmonic fit to the fractional difference of the CR rates from isotropic in equatorial coordinates as viewed by Milagro for the years 2000 - 2007. The color bin width is 1.0$\times$ 10$^{-4}$ reflecting the average statistical error. The two black lines show the position of the Galactic equator and the solid circle shows the position of the Galactic north pole. This map is constructed by combining 18 individual profiles of the anisotropy projection in R.A. of width 5$^{\circ}$ in decl. It is not a 2D map of the sky. The median energy of the events in this map is 6 TeV.
 \label{fig:milagro}}
  \end{center}
\end{figure}
%

The Milagro collaboration published in 2009 a 2D display of the sidereal anisotropy projections in R.A. at a primary CR energy of about 6 TeV \cite{milagro09}. The result is shown in the figure \ref{fig:milagro}. The map is  generated by combining 18 separate profiles of the anisotropy projection in R.A. of width 5$^{\circ}$ in declination. The measurement is in the R.A. direction only and no information in given about the R.A.-averaged CR rate difference from one declination band to another, i.e. on the anisotropy as a function of the declination band. 
In this sense, the figure \ref{fig:milagro} does not purport to be a picture of the anisotropy.

They observed a steady increase in the magnitude of the signal over seven years, in disagreement with the Tibet AS$\gamma$ results \cite{amenomori10}.
It is worth noting that the energy at which the Tibet AS$\gamma$ and Milagro results were obtained ($\sim$ 10 TeV) is too high for Sun effects play an important role.

\begin{figure}[!ht]
  \begin{center}
    \mbox{\epsfig{file=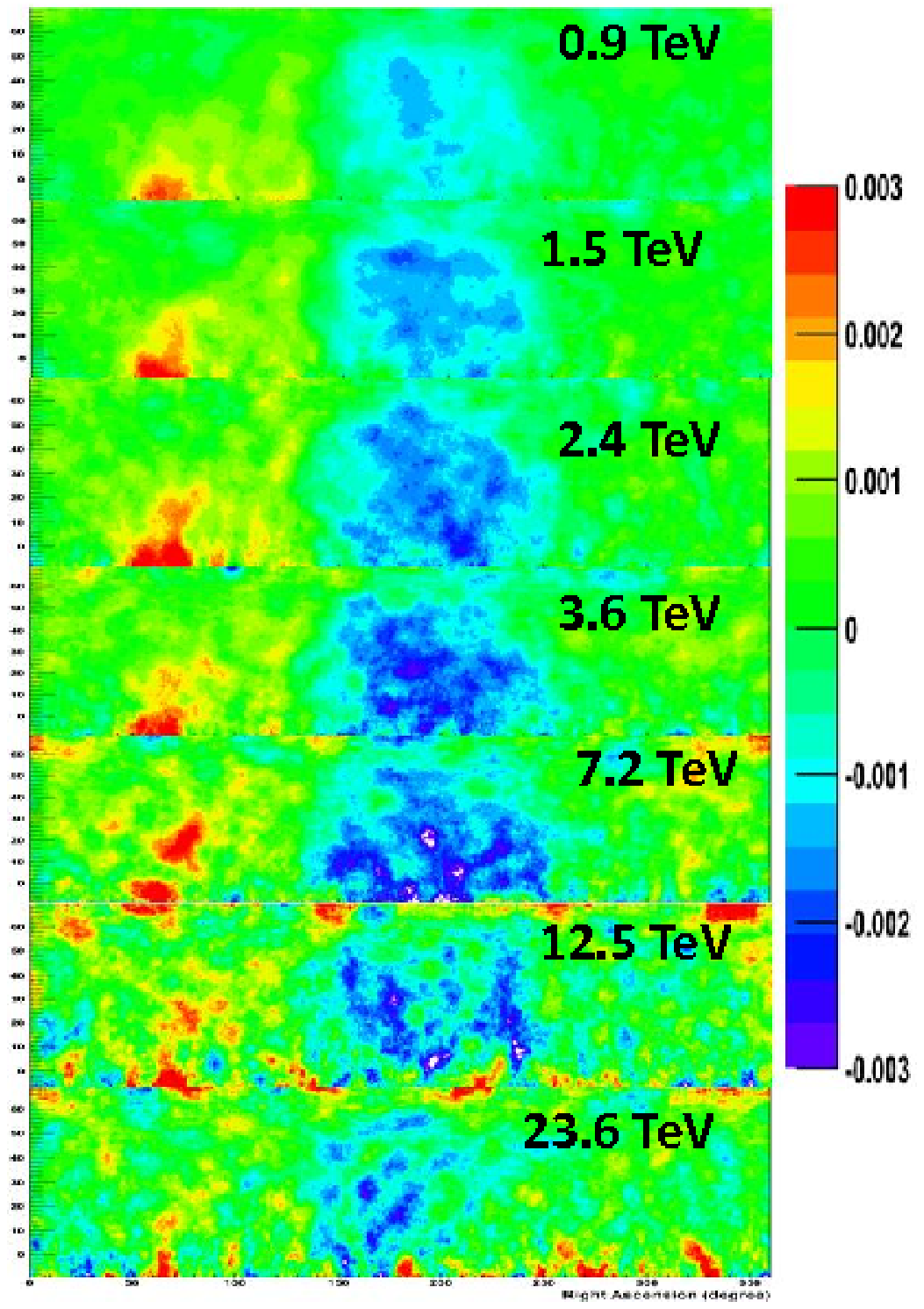,width=0.6\textwidth}}
  \vspace{-0.5pc}
    \caption{Large scale CR anisotropy observed by ARGO-YBJ as a function of the energy. The color scale gives the relative CR intensity.
 \label{fig:argo}}
  \end{center}
\end{figure}

The ARGO-YBJ experiment, located in Tibet (Yangbajing, 4300 m a.s.l.), reported the observation of 2D sky maps of large scale CR anisotropy in 2009 and 2011 for different primary energies \cite{argo09,argo11}. This is the first measurement of the CR anisotropy with an EAS array below the TeV energy region so far investigated only by underground muon detectors.
The results, shown in the figure \ref{fig:argo}, are in agreement with the findings of the Tibet AS$\gamma$ experiment, suggesting that the large anisotropy structures fade out to smaller spots when the energy increases.
%
%
%
\begin{figure}[!ht]
  \begin{center}
    \mbox{\epsfig{file=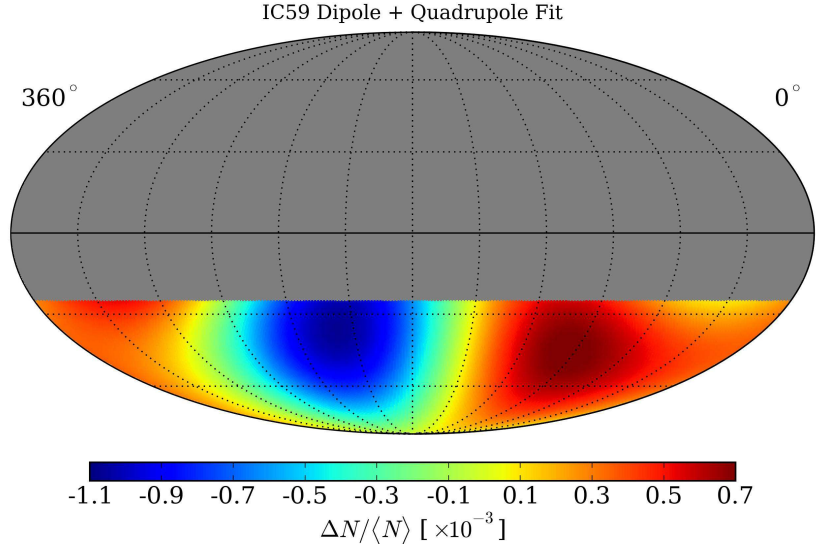,width=0.8\textwidth}}
  \vspace{-0.5pc}
  \caption{Sky distribution obtained by the IceCube experiment (IC59 configuration) by using the best-fit dipole and quadrupole coefficients.}
 \label{fig:icecube}
  \end{center}
\end{figure}
%

\begin{figure}[t]
\begin{tabular}{cc}
\subfigure[IceCube {\bf 20 TeV}]{\includegraphics[height=3.cm,width=0.47\textwidth]{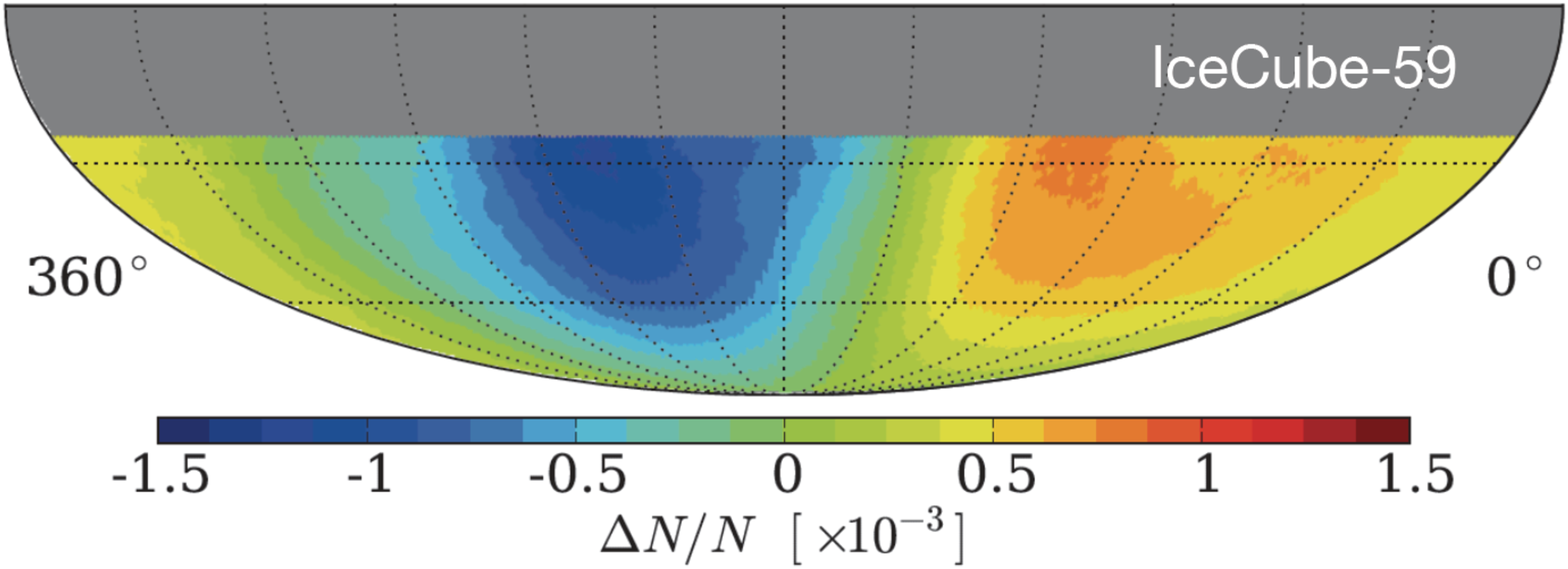} \label{fig:1mapa}} & 
\subfigure[IceCube {\bf 400 TeV}]{\includegraphics[height=3.cm,width=0.47\textwidth]{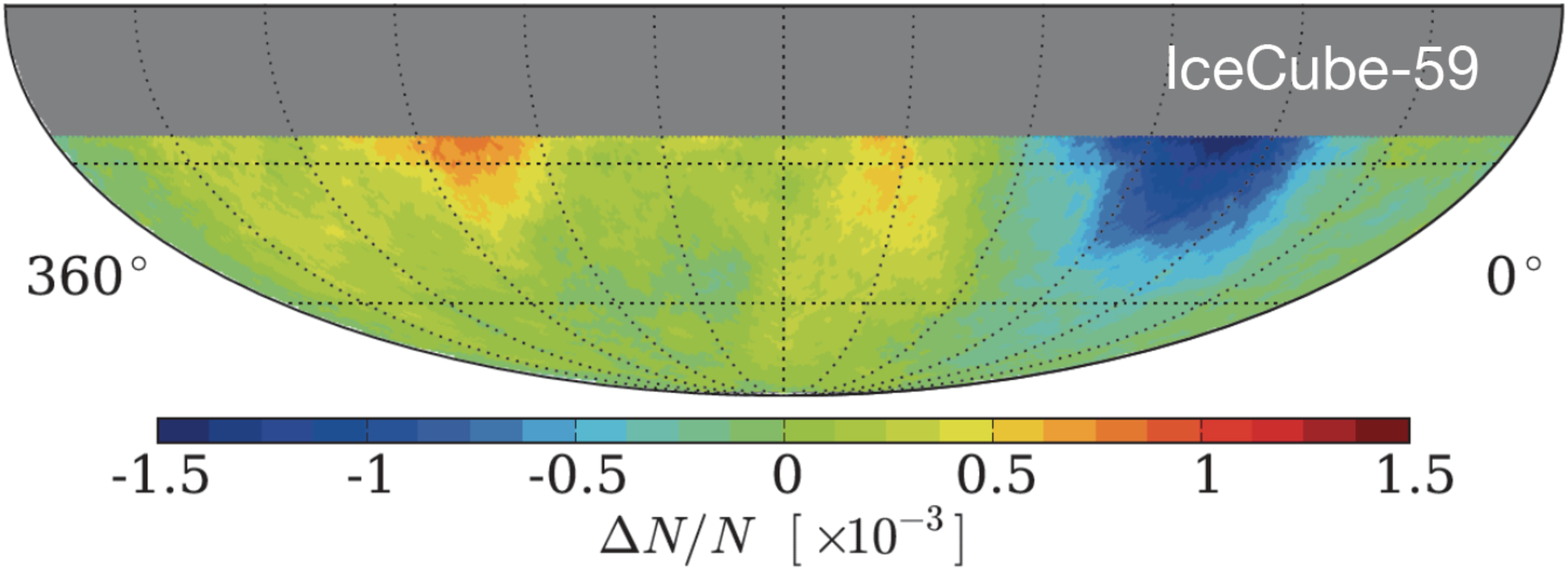} \label{fig:1mapb}} \\
\multicolumn{2}{r}{ } \\
\subfigure[IceTop {\bf 400 TeV}]{\includegraphics[height=3.cm,width=0.47\textwidth]{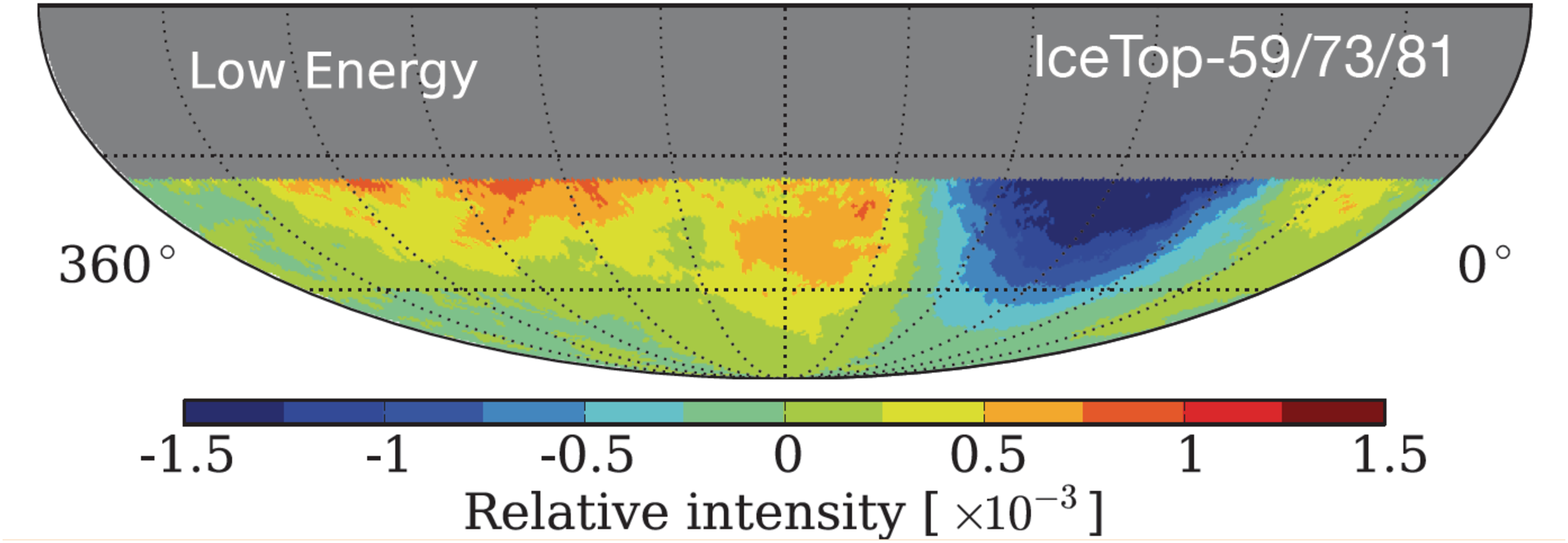} \label{fig:1mapc}} &
\subfigure[IceTop {\bf 2 PeV}]{\includegraphics[height=3.cm,width=0.47\textwidth]{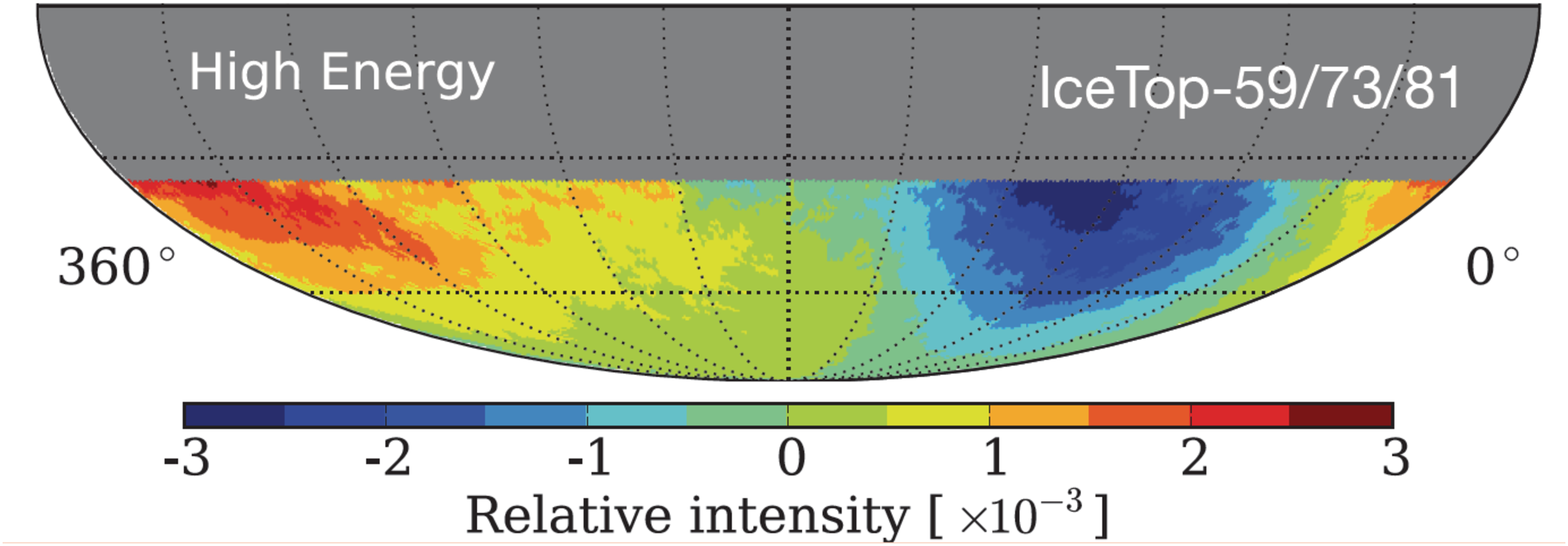} \label{fig:1mapd}} \\ 
\end{tabular}
\caption{2D maps of relative intensity in equatorial coordinates of the cosmic ray arrival distribution for IceCube (at 20 TeV and 400 TeV median energies) \cite{icecube12} and IceTop (at 400 TeV and 2 PeV median energies) \cite{itlarge13}.}
\label{fig:1maps}
\end{figure}

In 2010 the IceCube neutrino detector located at the South Pole reported the first 2D observation ever of the CR anisotropy in the southern hemisphere. This measurement was performed using muons generated in air showers by CRs with a median energy of 20 TeV \cite{icecube10}. 
Applying a spherical harmonic analysis to the relative intensity map of the CR flux, they observed significant anisotropy structures on several angular scales. Besides the large scale structure (see the figure \ref{fig:icecube}), in form of intense dipole
and quadrupole components, data showed several localized smaller features on scales between 15$^{\circ}$ and 30$^{\circ}$. The intensity of these spots is about a factor of five weaker than $\ell\leq$2 ones \cite{icecube11}. 

Another important result from the IceCube collaboration was the confirmation of the EAS-TOP finding, in the northern hemisphere, that a phase shift occurs around 400 TeV \cite{icecube12}, as can be seen in the Fig. \ref{fig:1mapa} and \ref{fig:1mapb}.
The figures show the map of relative intensity, in equatorial coordinates, of the arrival direction of cosmic rays observed by IceCube at median energies of about 20 TeV and 400 TeV, respectively. 
The anisotropy above 100 TeV was also measured using the surface EAS array IceTop at median energies of about 400 TeV and 2 PeV, as shown in Fig. \ref{fig:1mapc} and \ref{fig:1mapd}. The latter is the first determination of the anisotropy near the knee of the cosmic ray spectrum, showing the same global structure as at about 400 TeV, but with a deeper deficit \cite{itlarge13}.

The anisotropy at energies in excess of about 100 TeV has a different structure suggesting a possible transition in the contribution of nearby sources of cosmic rays \cite{blasi12, ptuskin1, ptuskin2}, or simply in the propagation mechanisms that generate the asymmetry \cite{desiati13}. 

Below about 100 TeV, the global anisotropy is dominated by the dipole and quadrupole components \cite{icecube11}. At higher energies the non-dipolar structure of the anisotropy challenges the current models of cosmic ray diffusion. Whether the strengthening of the deficit region with energy is due to propagation effects from a given source or to the contribution of heavier nuclei at the knee is not clear. In this regard, the study of cosmic ray spectrum and composition in correlation to the anisotropy and vice-versa of the anisotropy for different primary particle masses, would provide an experimental basis to test scenarios of particle transport in the local interstellar medium \cite{desiati-ricap13}.

\subsection{Recent Observations: Medium Scale Anisotropy}
\begin{figure}[!htbp]
  \begin{center}
    \mbox{\epsfig{file=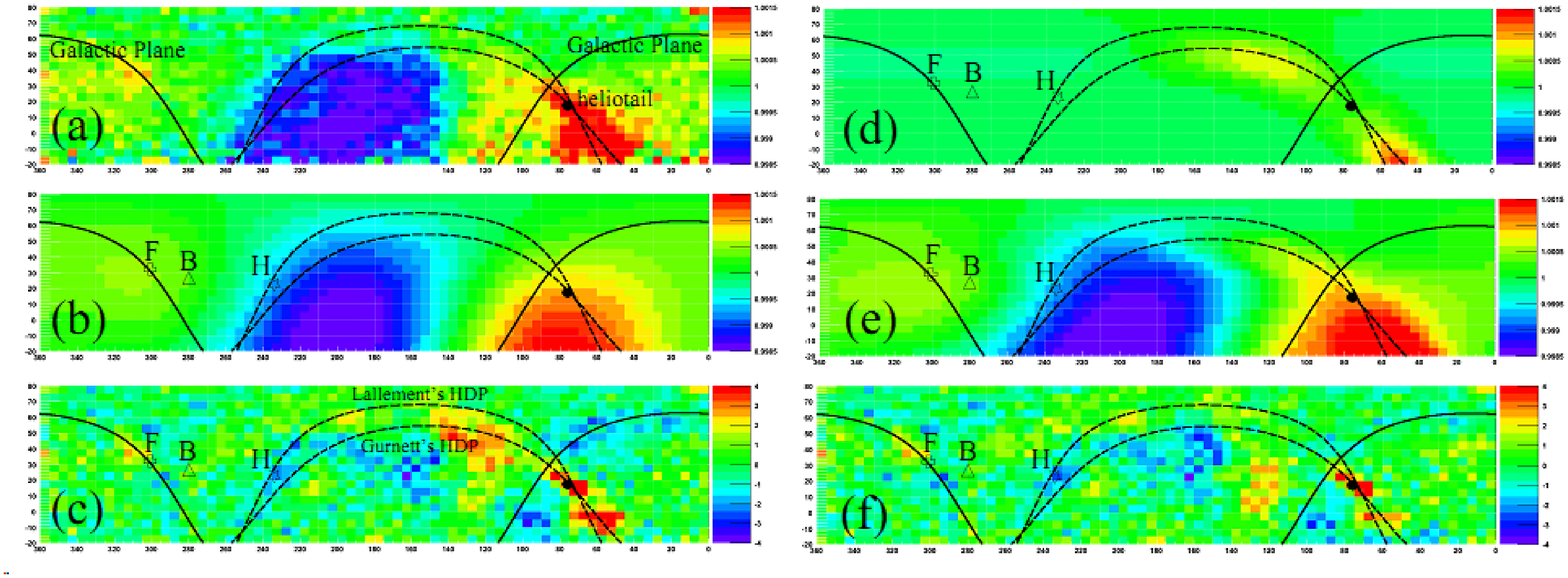,width=\textwidth}}
  \vspace{-0.5pc}
    \caption{2D anisotropy maps of galactic CRs observed and reproduced at the modal energy of 7 TeV by the Tibet-AS$\gamma$ experiment \cite{amenomori10}. (a): the observed CR intensity; (b): the best-fit large scale component; (c): the significance map of the residual anisotropy after subtracting the large scale component; (d): the best-fit medium scale component; (e): the best-fit large+medium scale components; (f): the significance map of the residual anisotropy after subtracting the large and the medium scale component. The solid black curves represent the galactic plane. The dashed black curves represent the Hydrogen Deflection Plane reported by \cite{gurnett06} and \cite{lallement05}. The helio-tail direction ($\alpha$, $\delta$) = (75.9$^{\circ}$ , 17.4$^{\circ}$) is indicated by the black filled circle. The open cross and the inverted star with the attached characters ``F'' and ``H'' represent the orientation of the local interstellar magnetic field by \cite{frisch96} and \cite{heerikhuisen10}, respectively. The open triangle with ``B'' indicates the orientation of the best-fit bi-directional cosmic-ray flow obtained in the reference \cite{amenomori10}.
 \label{fig:tibet_model}}
  \end{center}
\end{figure}

In 2007, modeling the large scale anisotropy of 5 TeV CR, the Tibet-AS$\gamma$ collaboration ran into a ``skewed'' feature over-imposed to the broad structure of the so-called tail-in region \cite{amenomori07,amenomori09}. They modeled it with a couple of intensity excesses in the hydrogen deflection plane \cite{gurnett06,lallement05}, each of them 10$^{\circ}$-30$^{\circ}$ wide. A residual excess remained in coincidence with the helio-tail. See the figure \ref{fig:tibet_model} (d) and its caption for more details.
%
%

\begin{figure}[!htbp]
  \begin{center}
    \mbox{\epsfig{file=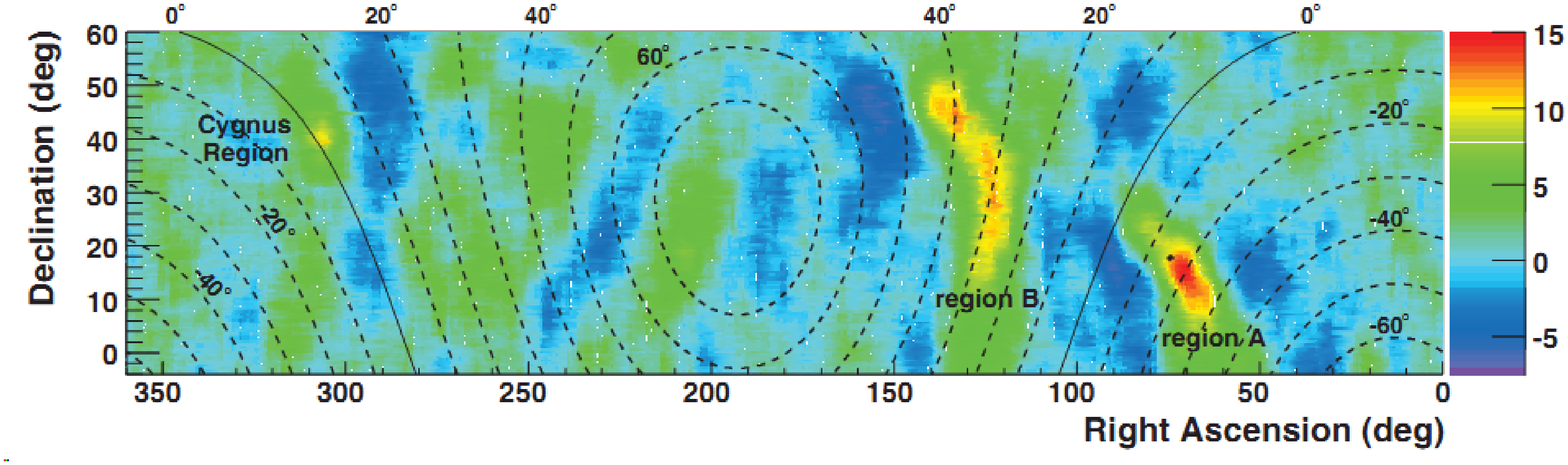,width=\textwidth}}
  \vspace{-0.5pc}
    \caption{Significance map for the Milagro data set without any cuts to remove the hadronic CR background. A 10$^{\circ}$ bin was used to smooth the data, and the color scale gives the statistical significance. The solid line marks the Galactic plane, and every 10$^{\circ}$ in Galactic latitude are shown by the dashed lines. The black dot marks the direction of the helio-tail, which is the direction opposite the motion of the solar system with respect to the local interstellar matter. 
 \label{fig:milagro2008}}
  \end{center}
\end{figure}

Afterwards the Milagro collaboration claimed the discovery of two localized regions of excess 10 TeV CRs on angular scales of 10$^{\circ}$ with greater than 12 $\sigma$ significance \cite{milagro2008}.
The figure \ref{fig:milagro2008} reports the pre-trial significance map of the observation. Regions ``A'' and ``B'', as they were named, are positionally consistent with the ``skewed feature'' observed by Tibet-AS$\gamma$.

The strongest and most localized of them (with an angular size of about 10$^{\circ}$) coincides with the direction of the helio-tail.
The fractional excess of region A is $\sim 6\times$10$^{-4}$, while for region B it is $\sim$ 4$\times$ 10$^{-4}$. The deep deficits bordering the excesses are due to a bias in the reference flux calculation. This effect slightly underestimates the significance of the detection.
The Milagro collaboration excluded the hypothesis of gamma-ray induced excesses.
In addition, they showed the excess over the large scale feature without any data handling (see the figure 2 of \cite{milagro2008}). 
%
%

The excesses in both regions are harder than the spectrum of the isotropic part of CRs. 

Easy to understand, more beamed the anisotropies and lower their energy, more difficult to fit the standard model of CRs and galactic magnetic field to experimental results. That is why these observations were rather surprising.
In addition, the observation of a possible small angular scale anisotropy region contained inside a larger one rely on the capability for suppressing the smooth global CR anisotropy at larger scales without, at the same time, introducing effects of the analysis on smaller scales.

Nonetheless, this observation has been confirmed by the ARGO-YBJ experiment in 2009 \cite{argomedium,argomedium2}. 

%
\begin{figure}
\centerline{\includegraphics[width=0.8\textwidth,clip]{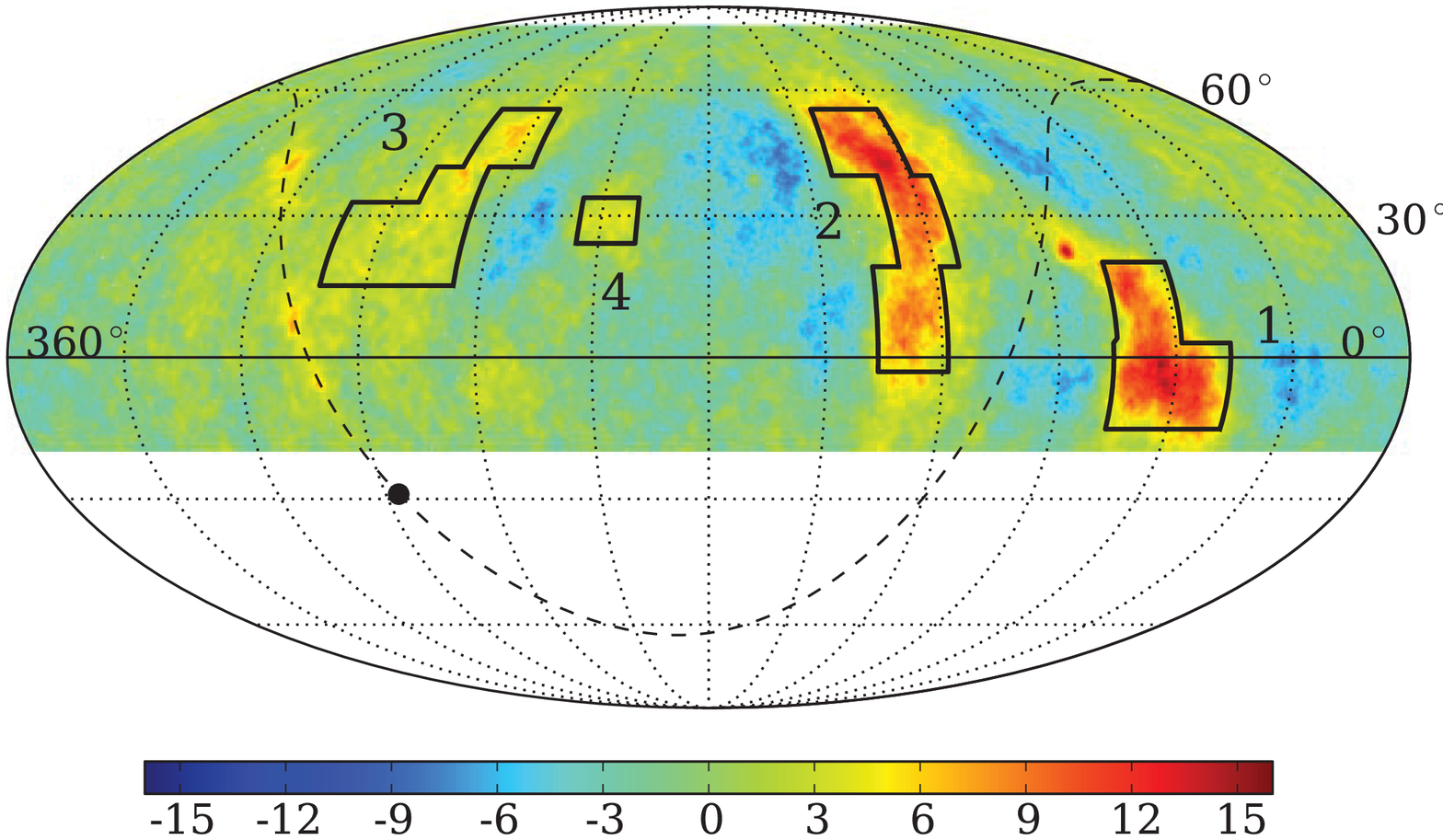} (a)}
\centerline{\includegraphics[width=0.8\textwidth,clip]{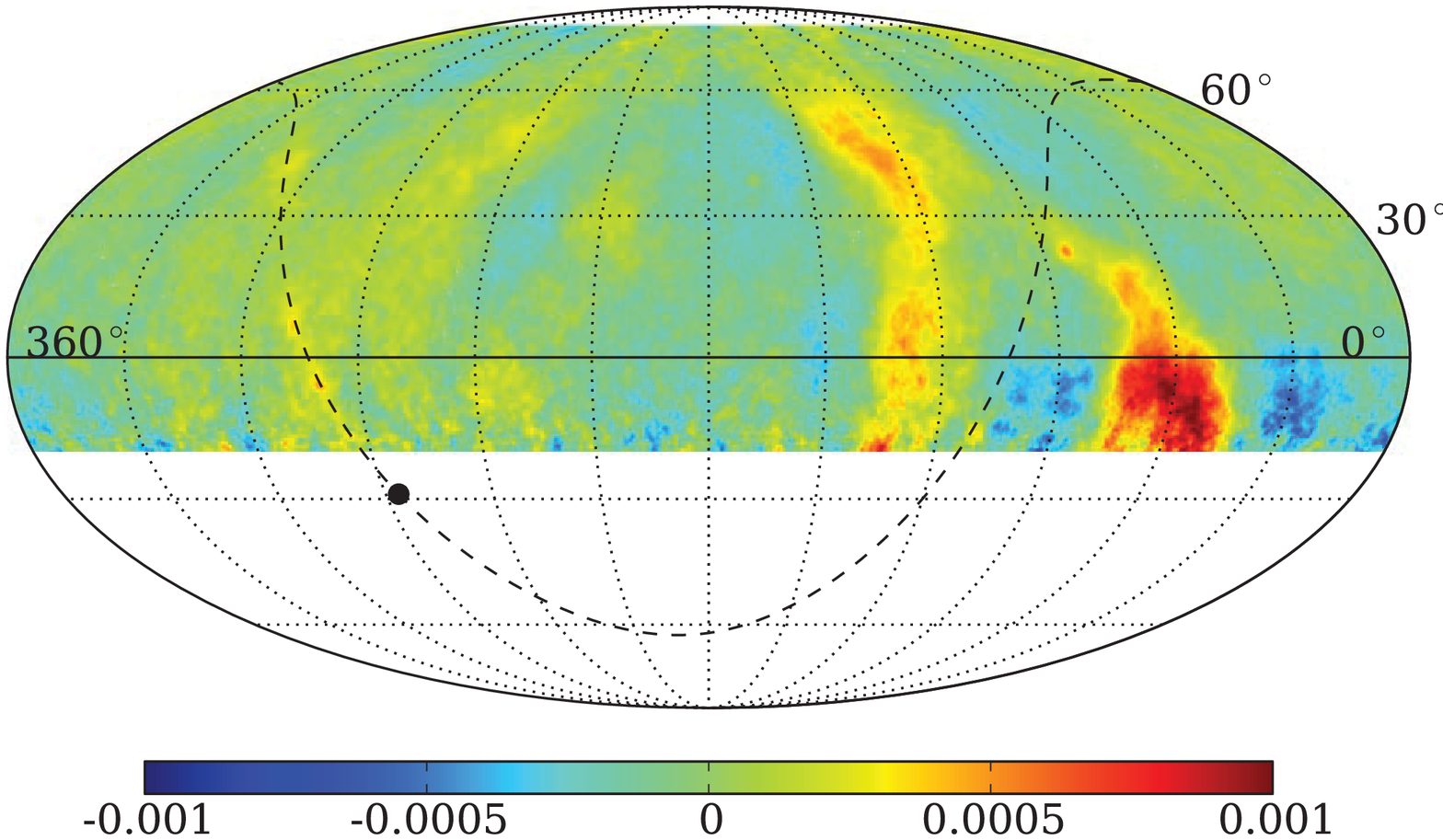} (b)}
\caption{ARGO-YBJ sky-map in equatorial coordinates for events with N$_{strip}\geq$ 25. The maps have been smoothed with an angle given by the PSF of the detector. {\em Plot (a):} statistical significance of the observation in s.d.. The boxes represent the parametrization of the regions of interest. {\em Plot (b):} relative excess with respect to the estimated background. The dashed line represents the Galactic Plane and the black point the Galactic Center.}
\label{fig:figA}       
\end{figure}

The Fig. \ref{fig:figA} shows the ARGO-YBJ sky map in equatorial coordinates as obtained with about 4$\cdot$10$^{11}$ events reconstructed with a zenith angle $\leq$50$^{\circ}$ \cite{bartoli13c}. According to the simulation, the median energy of the isotropic CR proton flux is E$_p^{50}\approx$1.8 TeV (mode energy $\approx$0.7 TeV).
No gamma/hadron discrimination algorithms have been applied to the data. Therefore, in the following the sky maps are filled with all CRs possibly including photons, without any discrimination.
The upper plot shows the statistical significance of the observation while the lower one shows the relative excess with respect to the estimated background.
They look slightly different because of the atmosphere thickness that the showers must cross before triggering the apparatus, increasing with the arrival zenith angle. As a consequence most significant regions do not necessarily coincide with most intense excesses. It should be noticed that also gamma-ray-induced signals are visible, because no gamma/hadron separation is applied.

%
\begin{figure}
\centerline{\includegraphics[width=0.8\textwidth,clip]{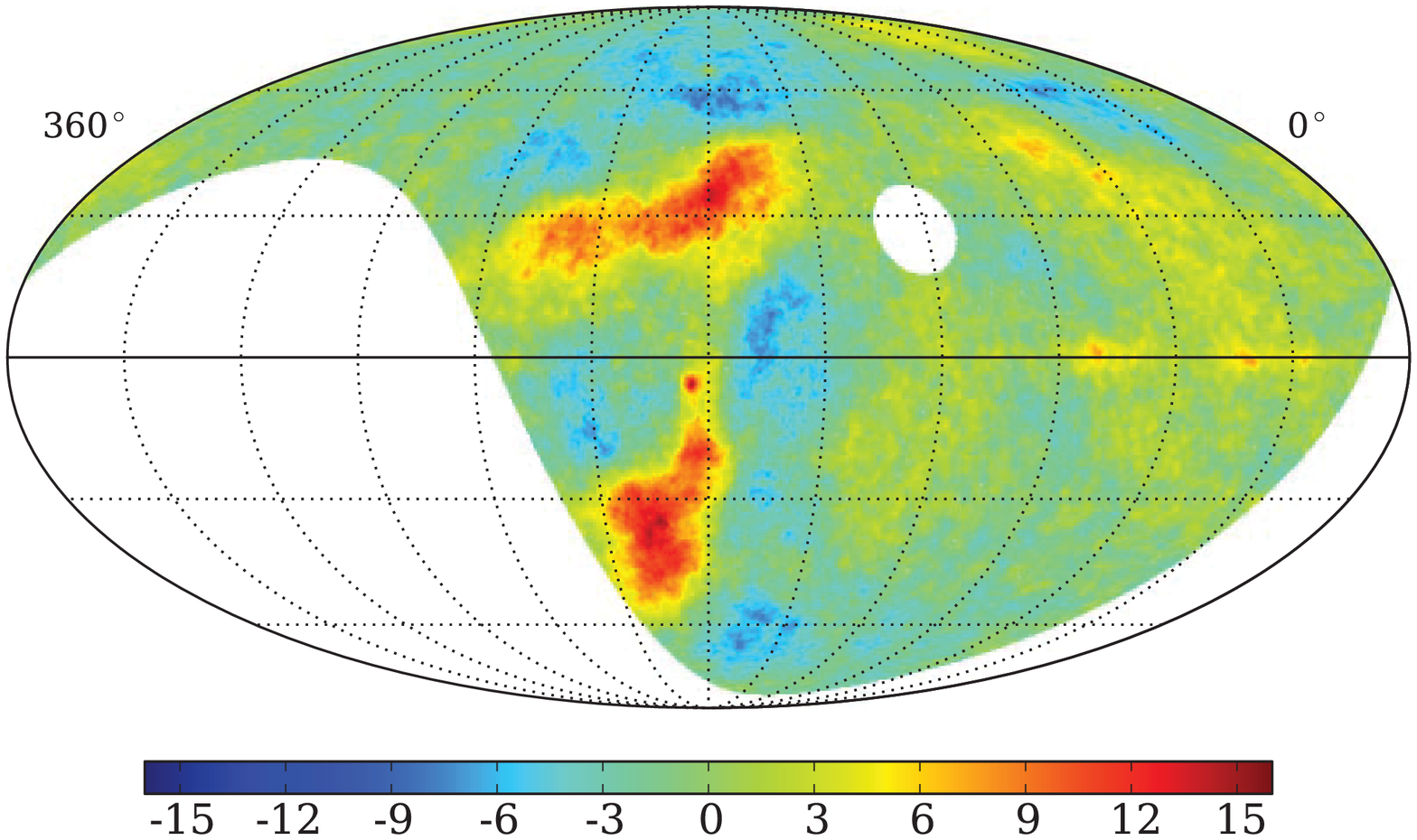}}
\caption{ARGO-YBJ sky-map of Fig. \ref{fig:figA}(a) in galactic coordinates. The map center points towards the anti-center.}
\label{fig:figD}       
\end{figure}
%
The most evident features are observed by ARGO-YBJ around the positions $\alpha\sim$ 120$^{\circ}$, $\delta\sim$ 40$^{\circ}$ and $\alpha\sim$ 60$^{\circ}$, $\delta\sim$ -5$^{\circ}$, spatially consistent with the regions detected by Milagro \cite{milagro2008}. These regions are observed with a statistical significance of about 15 s.d. and are represented on the significance map together with the other regions of interest.
As known from literature \cite{milagro2008,iuppadisc13}, the deficit regions parallel to the excesses are due to using also the excess events to evaluate the background, which turns out to be overestimated. Symmetrically, deficit regions, if any, would be expected to be surrounded by weaker excess halos, which were not observed.
On the left side of the sky map, several new extended features are visible, though less intense than the ones aforementioned. The area $195^{\circ}\leq R.A. \leq 290^{\circ}$ seems to be full of few-degree excesses not compatible with random fluctuations (the statistical significance is up to 7 s.d.). 
The observation of these structures is reported by ARGO-YBJ for the first time.

The upper plot of Fig. \ref{fig:figA} is represented in galactic coordinates in Fig. \ref{fig:figD}. 
In spite of the fact that the bulk of SNR, pulsars and other potential CR sources are in the Inner Galaxy surrounding the Galactic Centre, the excess of CRs is observed in the opposite, Anti-Centre direction. As remarked by Erlykin \& Wolfendale \cite{erlykin13}, the fact that the observed excesses are in the Northern and in the Southern Galactic Hemisphere, favors the conclusion that the CR at TeV energies originate in sources whose directions span a large range of Galactic latitudes.

In order to determine the energy of the four detected excesses, a suitable parameterization of their morphology has been introduced. For the sake of simplicity, all regions have been modeled with ``boxes'' in the (R.A., dec.) space. In case of complex shapes, a composition of boxes is used. The boxes are represented in the Fig. \ref{fig:figA} (a). They select the part of signal more than 3 s.d..

%
\begin{figure}
\centerline{\includegraphics[width=0.7\textwidth,clip]{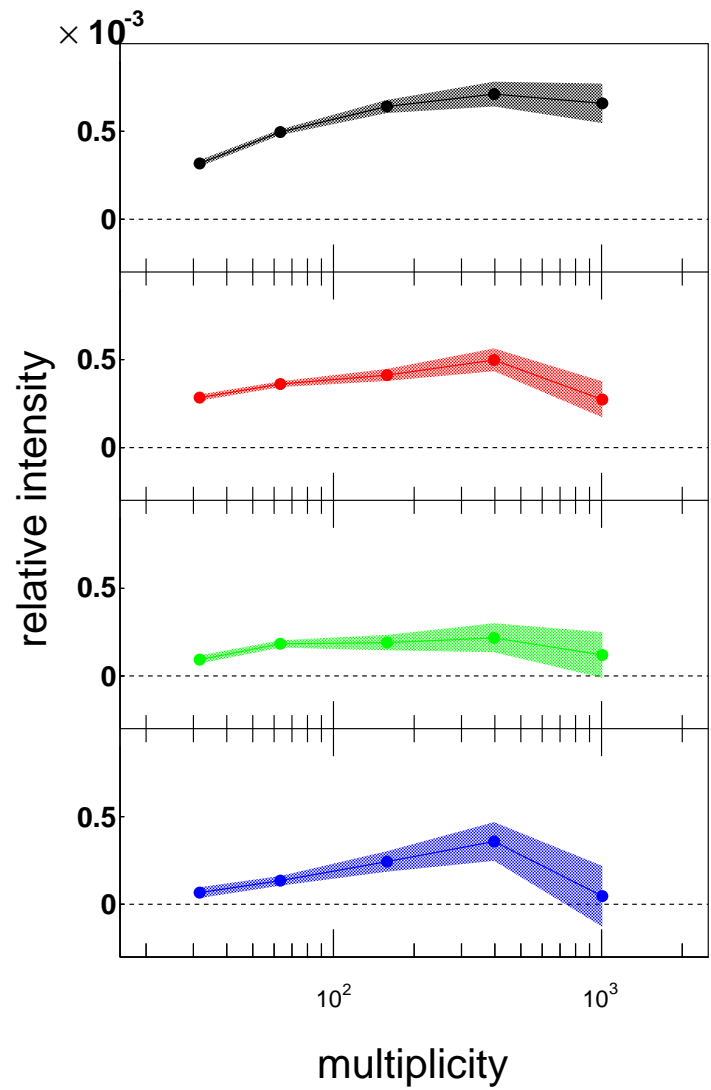}}
\caption{Size spectrum of the four MSA regions observed by ARGO-YBJ (regions 1 to 4 starting from the top). The vertical axis represents the relative excess $(e-b)/b$. The statistical errors are represented as coloured bands around the experimental points.}
\label{fig:energysp}       
\end{figure}

The Fig. \ref{fig:energysp} reports the multiplicity spectra for the anisotropy regions 1-4 (top-down). The number of events collected within each region are computed for the event map $e$ as well as for the background one $b$. The relative excess $(e-b)/b$ is computed for each multiplicity interval. The horizontal axis reports the multiplicity, the vertical one the relative intensity. 

The black plot reports the region 1 multiplicity spectrum. It is the hardest one detected by ARGO-YBJ and it shows a flattening around multiplicity 400 at relative intensity $\sim0.7\times10^{-3}$.
The region-2 multiplicity spectrum (red plot) is flatter than the one of region 1 and it turns out to be compatible with the constant result obtained by Milagro \cite{milagro2008}. The average intensity is $\sim0.35\times10^{-3}$.
Similar results are obtained for the region 3 (green graph), although the intensity is settled around $\sim0.2\times10^{-3}$.
The region 4 (blue graph), the least significant one, has a hard spectrum which rises up at a multiplicity between 300 and 400. 

Looking at the width of the error band (statistical error), it appears that for each region the multiplicity analysis gives a significant result up to N = 300--400. Instead, the high-multiplicity measurements are significant only for the regions 1 and 2, as the region 3 and 4 average excess is compatible with a null result.

The emission from region 1 is so intense and its observation so significant that interesting information can be obtained from the analysis of the multiplicity-energy relation in the sub-regions of its parametrization. In fact, the comparison of sub-region spectra is an important tool to check whether sub-regions are just geometrical parameterizations of the observed anisotropies, or they host different sources with various emission mechanisms.

%
\begin{figure}
\centerline{\includegraphics[width=0.7\textwidth,clip]{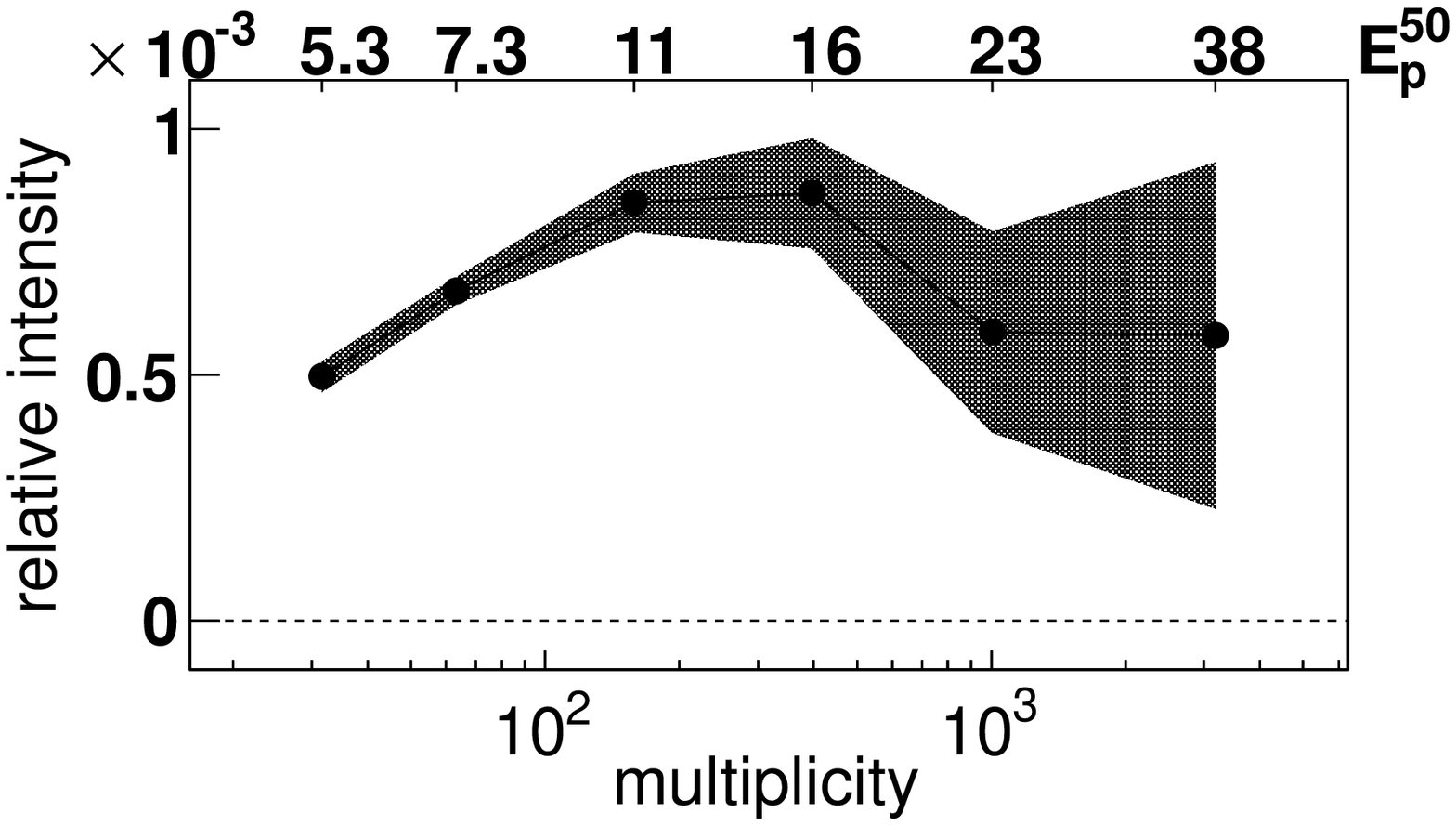}}
\centerline{\includegraphics[width=0.7\textwidth,clip]{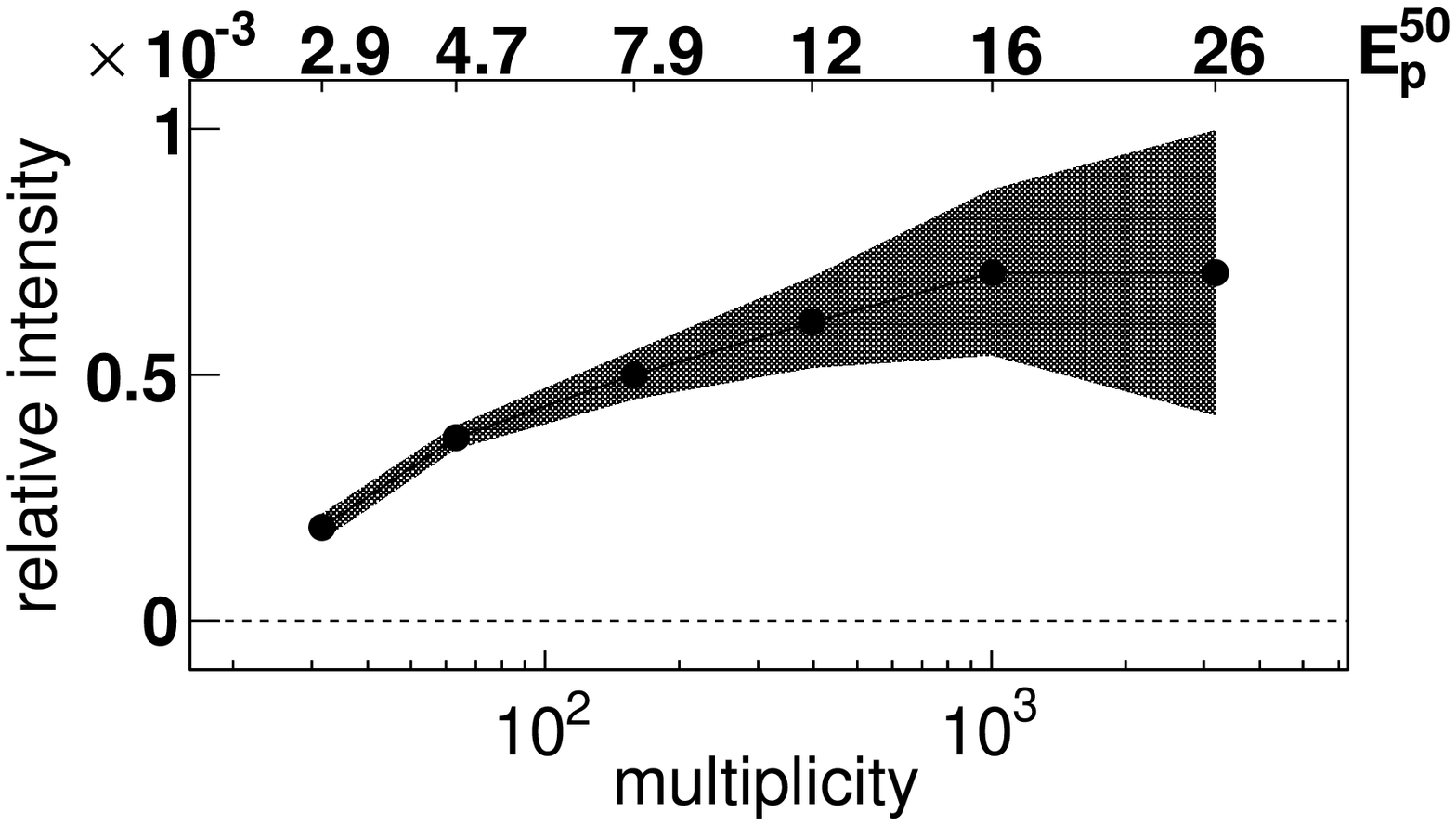} }
\caption{Multiplicity spectra of the sub-regions 1U (a) and 1L (b). The vertical axis represents the relative excess $(e-b)/b$. The upper horizontal scale shows the corresponding proton median energy (TeV).}
\label{fig:energysp-1}       
\end{figure}

The Fig. \ref{fig:energysp-1} poses the spectrum of the sub-regions 1U and 1L, with energy scales computed for a proton point-source having the average declination of each sub-region. The two regions are the upper and lower boxes of region 1 in the Fig. \ref{fig:figA}, respectively \cite{bartoli13c}.
To get more refined results at high energy, the last multiplicity bin (more than 630 fired strips) was split into $630-1599$ and $\geq 1600$.
For the region 1L a cut-off around 15--20 TeV can be noticed, compatible with that observed by Milagro in the region ``A'' \cite{milagro2008}. The statistics at high multiplicity is poor and does not allow to establish whether the cut-off continues at higher energy or not. Conversely, for region 1U a constantly increasing trend is obtained up to 26 TeV, what marks a possible difference between the sub-regions. Such a result, to be interpreted in the framework of a declination-dependent energy response, has to be compared with findings by higher energy experiments.

 As already said, the elemental composition and the energy spectrum are not known and that of CR protons is just an hypothesis. The ``photon'' hypothesis cannot be excluded \emph{a priori}, because in this work no gamma/hadron discrimination algorithms are applied. Even regions 1 and 2 exceed so much the Milagro parametrization that the conclusion about regions A and B not due to photons cannot be totally drawn.

Concerning the sub-parts of regions 2, 3 and 4, no significant features were found in their energy spectra, thus there is no reason to consider them more than just a simple geometrical parameterization.

\begin{figure}[!ht]
  \begin{center}
    \mbox{\epsfig{file=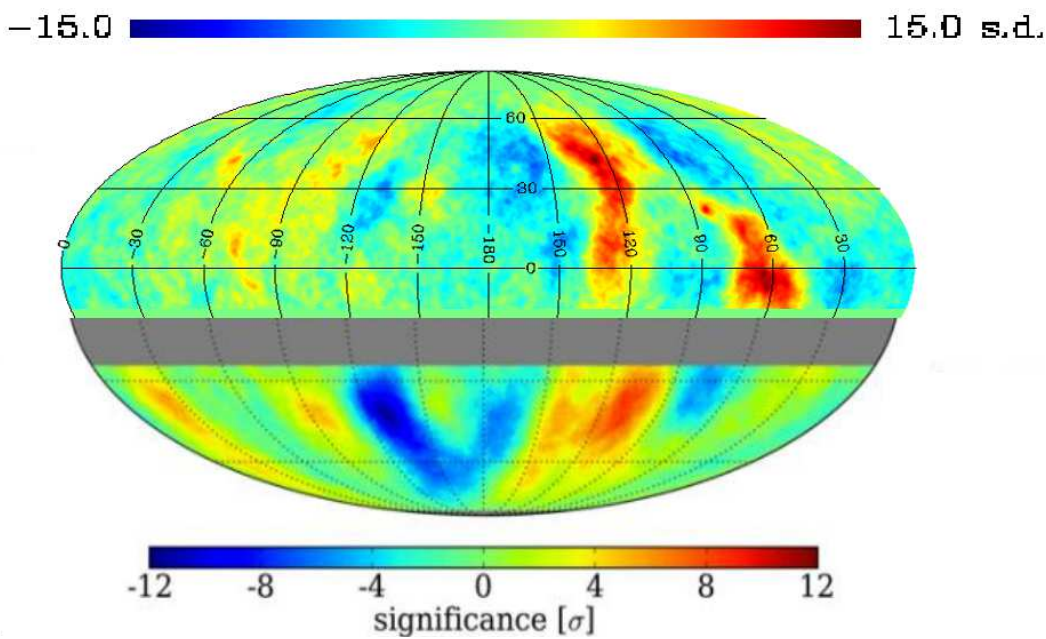,width=\textwidth}}
    \vspace{-0.5pc}
    \caption{Pictorial view of the CR TeV sky obtained by merging data from the ARGO.YBJ and the IceCube experiments. ARGO-YBJ covered the declination range $-20^{\circ}-80^{\circ}$, whereas the IceCube experiment observed the sky below $\delta=-65^{\circ}$. The image has illustrative purposes, as the median energy and the angular scale for which data were optimized are different for the experiments.}
      \label{fig:argoicecube}
  \end{center}
\end{figure}

Anticipated in the previous section, the IceCube collaboration found small scale features also in the Southern hemisphere \cite{icecube11}. In the figure \ref{fig:argoicecube} a pictorial view of the CR arrival distribution small scale structures all over the sky is reported. It has been realized for this paper by merging maps published by ARGO-YBJ and IceCube. 
%

It is worth recalling that the IceCube experiment measures muons, making us confident that charged CRs of energy above 10 TeV are observed.
%
\begin{figure}[!ht]
  \begin{center}
    \mbox{\epsfig{file=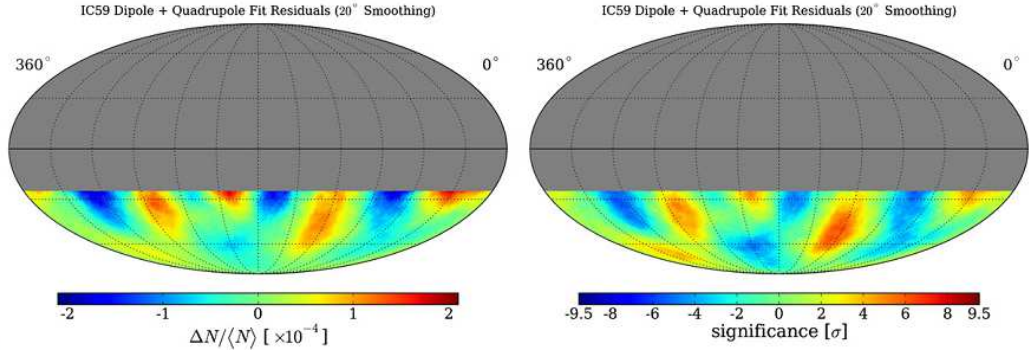,width=\textwidth}}
  \vspace{-0.5pc}
    \caption{Left: residual intensity map plotted with 20$^{\circ}$ smoothing. Right: significances of the residual map (pre-trials), plotted with 20$^{\circ}$ smoothing.
 \label{fig:icecubemed}}
  \end{center}
\end{figure}
%
%
\section{Models and Interpretations}
\label{sec:models}
Dealing with models of the results about the CR anisotropy, it should be noticed that people spent less effort for interpreting these observations than other traditional CR measurements, like the energy spectrum and the chemical composition. Perhaps, this is related to the few results available up to 1980s, as well as to the particular aspect of the anisotropy, which can be interpreted only if data are enough to cover wide portions of the sky. Surely, the problem of the anisotropy made a comeback with the last decade results and a number of theoretical models are expected to be formulated in the next future.

As discussed in the previous sections, on the basis of the sidereal diurnal modulations observed up to 100 TeV, most of the experiments reported a 24 hrs modulation as intense as 10$^{-4}$ - 10$^{-3}$ with phase of the maximum somewhere between 23 and 3 hr.

The origin of this anisotropy of galactic CRs is still unknown. 

In principle any anisotropy reflects a motion: (a) the motion of the Earth/Solar System with respect to the isotropic CRs rest frame, as suggested by Compton and Getting; (b) the motion of CRs from galactic sources to the extra-galactic space and/or from extra-galactic sources to the Earth. 

An observer moving with velocity $v$ relative to the rest frame of a CR plasma will detect a deviation due to this motion from the average CR intensity, an effect first described by Compton and Getting \cite{comptongetting35}. If the CRs have a differential power-law energy spectrum, E$^{-\gamma}$, then the fractional intensity enhancement due to the CG anisotropy is expressed as
\begin{equation}
\frac{\Delta I}{\langle I \rangle} = (\gamma + 2) \ \frac{v}{c}\ \cos\theta
\label{eq:cg}
\end{equation}
with $I$ denoting the CR intensity, $v/c$ the ratio of the detector velocity in the CR plasma rest frame and $\theta$ the angle between the observed CR and the moving direction of the detector \cite{gleeson68}. 
The eq. \ref{eq:cg} reads as a di-polar anisotropy with the maximum in the direction of the motion and a deficit in the opposite direction. 
We note that, for relativistic particles, the CG anisotropy does not depend on CR particle energy by equation \ref{eq:cg}. 

As it proceeds from a general principle, the CG effect is expected to generate observable anisotropy whenever the reference frame is suitably set up.
\begin{itemize}
\item A CG effect due to the Earth's motion around the Sun (Solar CG, SCG) is expected with an amplitude of about 0.05\% or less, depending on the geographic latitude of the detector, and with a maximum at 6:00 hr in the local solar time, as observed by different experiments;
\item if CRs did not co-rotate with the Galaxy, an observer on Earth would see an apparent excess of CR intensity towards the direction of galactic rotation and a deficit in the opposite one (Galactic CG, GCG).
The expected amplitude is of order 10$^{-3}$ with an apparent relative excess around 21 hr in sidereal time. 
\end{itemize}
The CG effect has been always considered as a benchmark for the reliability of the detector and the analysis method. In fact, at least as far as the SCG is concerned, all the features (period, amplitude and phase) of the signal are predictable without uncertainty, due to the exquisitely kinetic nature of the effect.
Actually, there may be some noise at low energy, due to the solar modulation and other possible local magnetic features not fully understood yet.

Nonetheless, above few TeV, where the solar modulation becomes negligible, several experiments measured the SCG effect in solar time and found it in agreement with the expectation. 

In 1986 using underground observations, Cutler and Groom reported the first clear signature of the SCG anisotropy for multi-TeV CRs observing a small diurnal amplitude modulation of the CR muon intensity \cite{cutler86}. The parent particles are sufficiently rigid ($\sim$ 1.5 TeV/c) that solar and geomagnetic effects should be negligible. 
The expected SCG at that latitude is 3.40$\times$10$^{-4}$, with the maximum at 06:00 local solar time. 
Analysis of the arrival times of 5$\times$ 10$^{8}$ muons during a period of 5.4 yr yields a fractional amplitude variation of 2.5$^{+0.7}_{-0.6}\times$10$^{-4}$, with a maximum at 08:18$\pm$1.0 h local mean solar time, deviating from 6:00 h by +2h at 2 $\sigma$ significance. They attributed the deviation to the meteorological effects on the underground muon intensity.

\begin{figure}[!ht]
  \begin{center}
    \mbox{\epsfig{file=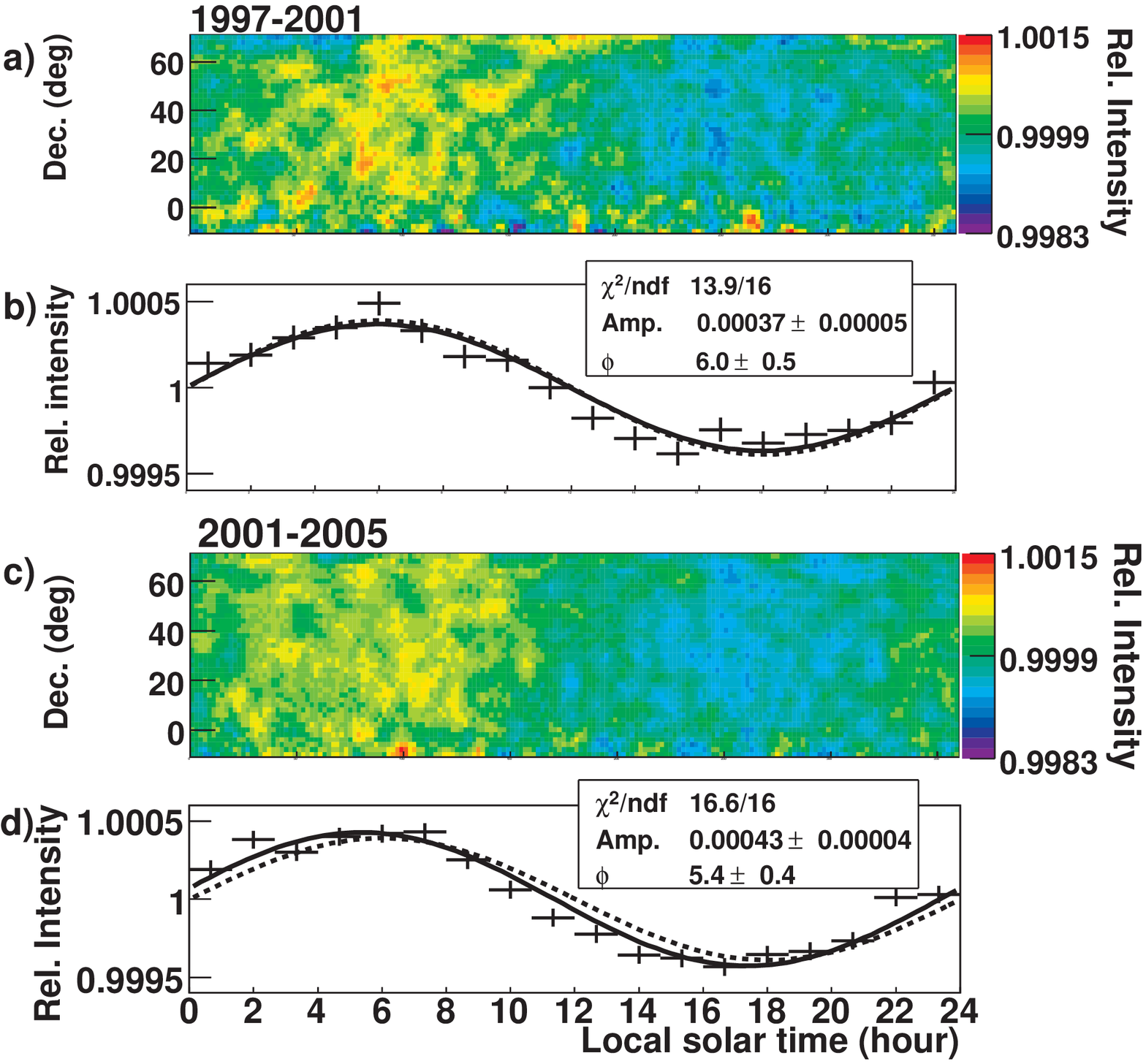,width=10.cm}}
  \vspace{-0.5pc}
 \caption{Solar time anisotropy observed by the Tibet AS$\gamma$ experiment in 2D at modal energy of 10 TeV \cite{amenomori06}. Figures a) and b) are the 2D map and the 1D projection of data collected in 1997-2001. Figures c) and d) refer to the period 2001-2005. The black solid line is the best-fit sine function, whereas the dashed line is the expected CG effect.
 \label{fig:cg_tibet}}
  \end{center}
%
  \begin{center}
    \mbox{\epsfig{file=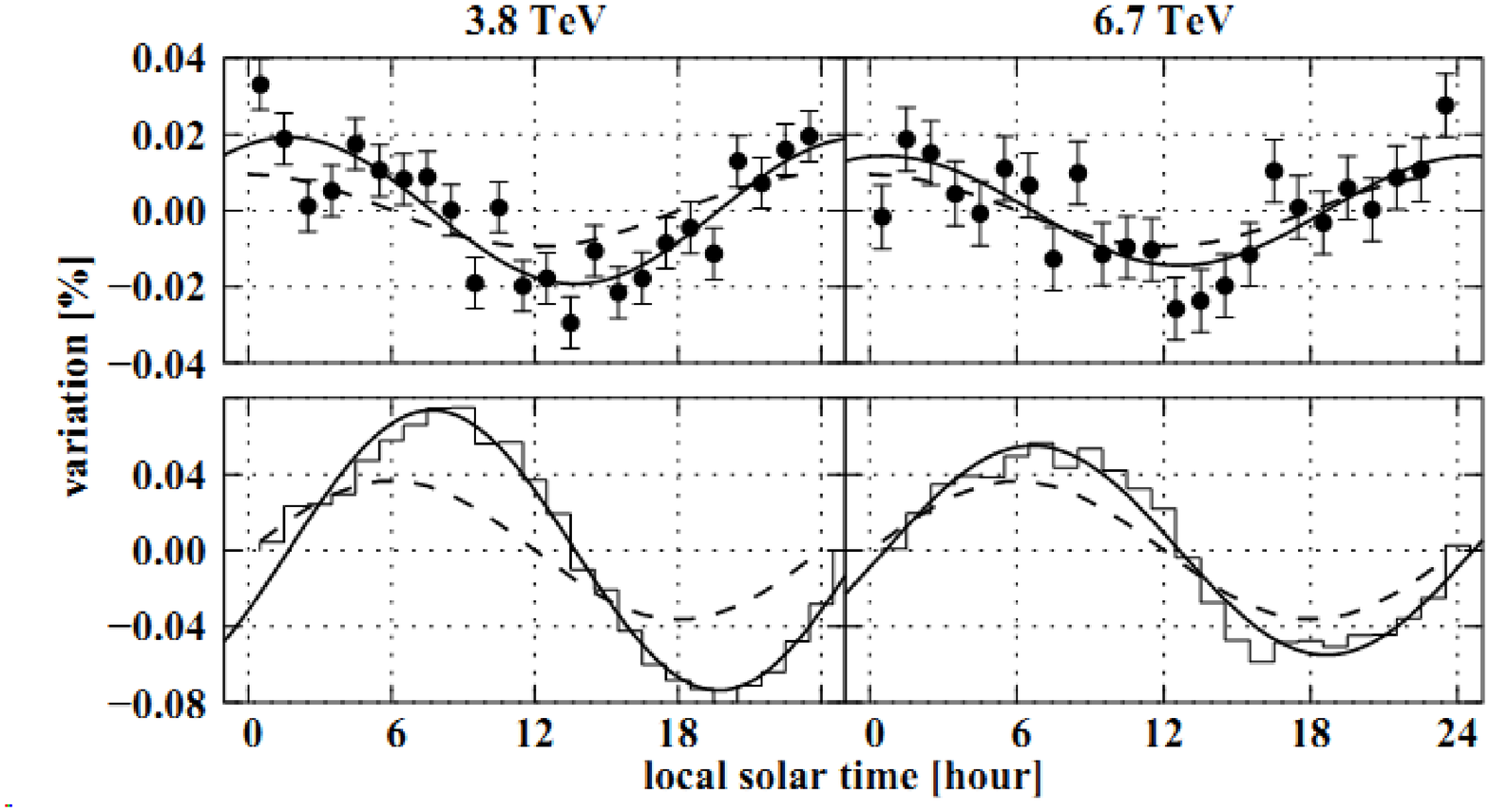,width=10.cm}}
  \vspace{-0.5pc}
  \caption{Observed solar daily variations divided into two energy regions compared with the expected anisotropy due to the CG effect (the broken lines). The plot in the upper panel is the differential form of the solar daily variations, and histogram in the lower panel is physical daily variations. Solid curves are the two-parameter $\chi^2$-fitted sine curves.
 \label{fig:cg_tibet_ener}}
    \end{center}
 \end{figure}

The first clear observation of the SCG effect with an EAS array was reported by the EAS-TOP collaboration in 1996 with a statistical significance of 7.3 $\sigma$ at an energy of about 10$^{14}$ eV \cite{aglietta96}.
A detailed study of the SCG effect has been carried out by the Tibet AS$\gamma$ experiment. In the figure \ref{fig:cg_tibet} the solar time anisotropy observed in two different periods is shown. In the figure \ref{fig:cg_tibet_ener} the anisotropy is reported for two different primary CR energies. 
The modulations show that the observations follow the SCG model at 6.7 TeV (right panel), but some deviations are there at lower energy, 3.8 TeV (left panel). This finding suggested that some related effects may influence the CR arrival direction distribution also in the few TeV energy region \cite{amenomori03}.
%
%

In spite of these successful results about the SCG, the arrival distribution in sidereal time was never found to be purely di-polar, neither any signature of the GCG effect was ever observed in sidereal time. 

Consequently, the CR plasma is supposed to co-move with the solar system and the origin of the observed anisotropy is thought to be related to ``harder'' effects, to be searched for in unknown features of the local interstellar medium (LISM), either for the magnetic field and the closest CR sources. 

Some authors suggested that the large scale anisotropy can be explained within the diffusion approximation taking into account the role of the few most nearby and recent sources \cite{blasi12,erlykin06,giacinti11}.
Other studies suggest that a non-di-polar anisotropy could be due to a combined effect of the regular and turbulent GMF \cite{battaner09}, or to local uni- and bi-dimensional inflows \cite{amenomori10}. In particular the authors modeled the observed anisotropy by a superposition of a large, global anisotropy and a midscale one. The first one is proposed to be generated by galactic CRs interacting with the magnetic field in the local interstellar space surrounding the heliosphere (scale $\sim$2 pc). The midscale anisotropy is possibly caused by a modulation of galactic CRs in the helio-tail \cite{amenomori11}.

The highest energy observations by EAS-TOP and IceCube \cite{aglietta09,icecube12} of the existence of a new anisotropy around 400 TeV suggest that the global anisotropy is the superposition of different contributions due to phenomenologies at different distances from the Earth \cite{amenomori10,desiati10}. 
A possible underlying anisotropy due to the CR sources distribution seems to manifest itself at energies above 100 TeV, being at lower energies the proton gyro-radius of the same order of the helio-tail \cite{pogorelov09a,pogorelov09b}.
Therefore, probably in the one hundred TeV energy region we are observing a transition in the cause of the anisotropy, being mainly due to local solar effects the lower energy phenomenology.

About the medium scale anisotropy, no theory of CRs in the Galaxy exists yet which is able to explain few degrees anisotropies in the rigidity region 1-10 TV leaving the standard model of CRs and that of the local galactic magnetic field unchanged at the same time. 
First interpretations based on observing the excess are inside the ``tail-in'' zone, that induced authors to drag in interactions of CRs with the heliosphere\cite{milagro2008}.
Several authors, noticing that the TeV region is usually free of heliosphere-induced effects, proposed a model where the excesses are produced in the Geminga supernova explosion \cite{SalvatiSacco}. In the first variant of the model CRs simply diffuse (Bohm regime) up to the solar system, while the second version limits the diffusion to the very first phase of the process and appeals to non-standard diverging magnetic field structure to bring them to the Earth. 
Other people \cite{DruryAharonian} proposed similar schemes involving local sources and magnetic traps guiding CRs to the Earth. It must be noticed that sources are always intended to be near-by, at less than 100 - 200 pc. Moreover the position of the excesses in galactic coordinates, symmetrical with respect to the galactic plane, played an important role in inspiring such models.

Grounding on the observation that all nearby sources and new magnetic structures brought in to explain the medium scale anisotropy should imply other experimental signatures that would have been observed in the past, some other  models were proposed. 
In \cite{amenomori10,amenomori11} the authors suggest that the magnetic field in the helio-tail (that is, within $\sim$70 AU to $\sim$340 AU from the Sun) is responsible for the observed midscale anisotropy in the energy range of 1 - 30 TeV. In particular, this effect is expressed as two intensity enhancements placed along the hydrogen deflection plane, each symmetrically centered away from the helio-tail direction.

The hypothesis that the effect could be related to the interaction of isotropic CRs with the heliosphere has been re-proposed in \cite{desiati10}. Grounding on the coincidence of the most significant localized regions with the helio-spheric tail, magnetic reconnection in the magneto-tail has been shown to account for beaming particles up to TeV energies. 
On a similar line, in \cite{desiati11} the authors suggest that even the large scale anisotropy below 100 TeV is mostly due to particle interactions with the turbulent ripples generated by the interaction of the helio-spheric and interstellar magnetic field. This interaction could be the dominant factor that re-distribute the large scale anisotropy from an underlying existent anisotropy. At the same time, such scattering can produce small angular scale features along line of sights that are almost perpendicular to the local interstellar magnetic fields.

Recently, the role of interactions of TeV - PeV CRs with a turbulent magnetic field across they propagate through has been emphasized to explain the small scale anisotropy \cite{giacinti11}. The authors show that energy-dependent medium and small scale anisotropies necessarily appear, provided that there exists a large scale di-polar anisotropy, for instance from the inhomogeneous source distribution. The small scale anisotropies naturally arise from the structure of the local turbulent GMF, typically within the CR scattering length.

There has been also the suggestion that CRs might be scattered by strongly anisotropic Alfven waves originating from turbulence across the local field direction \cite{MalkovEtAl}.

Besides all these ``ad hoc'' interpretations, several attempts occurred in trying to insert the CR excesses in the framework of recent discoveries from satellite-borne experiments, mostly as far as leptons are concerned. In principle there is no objection in stating that few-degree CR anisotropies are related to the positron excess observed by Pamela \cite{PamelaPositrons} and to the electrons excess observed by Fermi \cite{FermiPositrons}. All observations can be looked at as different signatures of common underlying physical phenomena.
\section{Conclusions}
The cosmic ray arrival direction distribution and its anisotropy has been a long-standing problem ever since the 1930s. In fact, the study of the anisotropy is a powerful tool to investigate the acceleration and propagation mechanism determining the CR world as we know it. Nonetheless, with respect to other classics of CR physics, fewer experiments managed to get significant results, due to the small intensity of the effect.

From the theoretical viewpoint, apart from some effects of kinetic nature (SCG and GCG effect), no signal is expected either in the sidereal and the solar time reference frame. If the energy is high enough, i.e. solar effects are negligible, the standard model of production, acceleration and propagation of CRs does not foresee any deviation from the isotropy. In this picture of the outer space, a major role is played by the magnetic fields the CRs pass through before touching the Earth atmosphere. When the energy increases enough, a weak dipole may arise, as a mild signature of the closest and most recent CR sources.

So far, the anisotropy in the CR arrival direction distribution have been observed by different experiments with increasing sensitivity and details at different angular scales.

Current experimental results show that the main features of the anisotropy are uniform in the energy range (10$^{11}$ - 10$^{14}$ eV), both with respect to amplitude (10$^{-4}$ - 10$^{-3}$) and phase ((0 - 4) hr).
The existence of two distinct broad regions, one showing an excess of CRs (called ``tail-in''), distributed around 40$^{\circ}$ to 90$^{\circ}$ in R.A., the other a deficit (the ``loss cone''), distributed around 150$^{\circ}$ to 240$^{\circ}$ in R.A., has been clearly observed.

The framework became clearer in the last decade, when more and more experiments demonstrated to have the capability of drawing 2D sky-maps, thus giving to the scientific community the evidence of localized regions whose physical nature is still unknown.

In the last years the Milagro and ARGO-YBJ Collaborations reported evidence of the existence of a medium angular scale anisotropy contained in the tail-in region. The observation of similar small scale anisotropies has been recently claimed also by the IceCube experiment in the southern hemisphere.

So far, no theory of CRs in the Galaxy exists which is able to explain both large scale and few degrees anisotropies leaving the standard model of CRs and that of the local galactic magnetic field unchanged at the same time. 

A joint analysis of concurrent data recorded by different experiments in both hemispheres, as well as a correlation with other observables like the interstellar energetic neutral atoms distribution \cite{ibex09,ibex11}, should be a high priority to clarify the observations.

 
~

MA

\label{lastpage-01}
\end{document}